\documentclass[reprint,aps,prb,english,nolongbibliography,superscriptaddress]{revtex4-2}

\usepackage{graphicx}
\usepackage{bm,amssymb,amsmath,mathrsfs,amstext,latexsym,physics}
\usepackage[T1]{fontenc}

\usepackage[dvipsnames]{xcolor}
\usepackage[colorlinks=true,citecolor=blue,urlcolor=blue,linkcolor=blue]{hyperref}
\usepackage[normalem]{ulem}

\def\rmd{\mathrm{d}}

\begin{document}

\title{Scaling analysis and renormalization group on the mobility edge\\in the quantum random energy model}

\author{Federico Balducci}
\email{fbalducci@pks.mpg.de}
\affiliation{Max Planck Institute for the Physics of Complex Systems, N\"othnitzer Str. 38, 01187 Dresden, Germany}

\author{Giacomo Bracci Testasecca}
\email{gbraccit@sissa.it}
\affiliation{SISSA, via Bonomea 265, 34136, Trieste, Italy}
\affiliation{INFN, Sezione di Trieste, Via Valerio 2, 34127 Trieste, Italy}

\author{Jacopo Niedda}
\email{jniedda@ictp.it}
\affiliation{The Abdus Salam ICTP, Strada Costiera 11, 34151 Trieste, Italy}

\author{Antonello Scardicchio}
\affiliation{The Abdus Salam ICTP, Strada Costiera 11, 34151 Trieste, Italy}
\affiliation{INFN, Sezione di Trieste, Via Valerio 2, 34127 Trieste, Italy}

\author{Carlo Vanoni}
\email{cv9865@princeton.edu}
\affiliation{Department of Physics, Princeton University, Princeton, New Jersey, 08544, USA}

\date{\today}


\begin{abstract}
    Building on recent progress in the study of Anderson and many-body localization via the renormalization group (RG), we examine the scaling theory of localization in the quantum Random Energy Model (QREM). The QREM is known to undergo a localization-delocalization transition at finite energy density, while remaining fully ergodic at the center of the spectrum. At zero energy density, we show that RG trajectories consistently flow toward the ergodic phase, and are characterized by an unconventional scaling of the fractal dimension near the ergodic fixed point. When the disorder amplitude is rescaled, as suggested by the forward scattering approximation approach, a localization transition emerges also at the center of the spectrum, with properties analogous to the Anderson transition on expander graphs. At finite energy density, a localization transition takes place without disorder rescaling, and yet it exhibits a scaling behavior analogous to the one observed on expander graphs. The universality class of the model remains unchanged under the rescaling of the disorder, reflecting the independence of the RG from microscopic details. Our findings demonstrate the robustness of the scaling behavior of random graphs and offer new insights into the many-body localization transition.
\end{abstract}

\maketitle

\section{Introduction} 

Since the early days of statistical mechanics, physicists have been interested in understanding how the collective behavior of many degrees of freedom leads to the macroscopically observed phenomena. While sometimes the thermodynamic limit can be accessed more or less straightforwardly, in many other situations finite-size corrections are very strong, resulting from interesting and non-trivial physical processes taking place on different scales. This is particularly true in the context of quantum localization transitions, arising from the competition between tunneling and disorder~\cite{Anderson1958absence,evers2008anderson}. Already in the single-particle setting, extensive numerical simulations have been required to extract reliable estimates of the critical exponents~\cite{Slevin-PRL99,evers2008anderson,ueoka2014dimensional,Tarquini2017critical}. Upon introducing interactions, the very existence of a stable many-body localized (MBL) phase has been debated for over 40 years~\cite{fleishman1980interactions,altshuler1997quasiparticle,Gornyi2005Interacting,Basko06,oganesyan2007localization,Pal10,Bardarson2012Unbounded,Serbyn2013Universal,*Serbyn2013Local,Diggen2018Manybody,Deluca13,Suntajs2020Quantum,Sels2021Dynamical}, and only recently rigorous proofs have settled some aspects of the debate~\cite{Imbrie2016Diagonalization,*Imbrie2016Many,deroeck2024absence}. A great deal of controversy on the topic is due to the presence of severe finite-size effects~\cite{Chandran2015Finite,panda2020can,abanin2021distinguishing,sierant2022challenges,sierant2023stability,sierant24MBLreview}, masking the thermodynamic limit in numerical studies, and possibly giving rise to intermediate regimes~\cite{Luitz2017Ergodic,Crowley2022Constructive,Morningstar2022Avalanches}. In turn, analytical control for intermediate system sizes is poor because a reliable scaling theory of localization~\cite{abrahams1979scaling} in the presence of interactions is lacking, as the relevant---in the renormalization group (RG) sense---fields driving the transition have to be identified with confidence yet~\cite{Vosk2013Many,Potter2015Universal,Zhang2016Many,Thiery2017Microscopically,Goremykina2019Analytically,Dumitrescu2019Kosterlitz,Morningstar2019Renormalization,Morningstar2020Many}.

Most of the analytical understanding of localization transitions is based on perturbative expansions. In the many-body case, the expansions are performed for effective single-particle models on Fock space graphs, under the assumption that loops have a negligible effect~\cite{Basko06,ros2015integrals}. For this reason, the Anderson model on expander graphs---such as random regular graphs (RRG)---has been argued to capture some of the phenomenology of the many-body case, and it has been studied extensively~\cite{de2014anderson,tikhonov2016Anderson,Tarquini2017critical,bera2018return,kravtsov2018non,parisi2019anderson,tikhonov2021AndersonMBL,sierant2023universality,vanoni2023analysis}. Recently, however, a RG approach that makes use of modern numerical techniques has shed new light on the problem~\cite{vanoni2023renormalization}. A two-parameter scaling (2PS) hypothesis has been shown to describe the emergence of the localized phase, which corresponds to a line of fixed points, reminiscent of a Berezinskii-Kosterlitz-Thouless (BKT) scenario~\cite{Berezinskii1971Destruction,*Berezinskii1972Destruction,Kosterlitz1973Ordering}. This is in contrast with the transition in finite dimensions, where the fixed points of the transition, of the localized phase and of the ergodic phase are all described by a one-parameter scaling (1PS) RG flow~\cite{abrahams1979scaling,altshuler2024renormalization}.

In this work, we take a step further and analyze the dynamical phase diagram of the quantum random energy model (QREM)~\cite{Laumann2014MBMobility,Baldwin2016ManyBody} via the RG approach of Refs.~\cite{vanoni2023renormalization,kutlin2024investigating,altshuler2024renormalization,Niedda2024,Swietek_2024}. The QREM is a toy model for many-body localization transitions, as it retains the Fock space structure of many-body hopping, while disregarding the correlations that build up in the disorder term. The QREM can also be regarded as an unconventional infinite-dimensional limit of the Anderson model, where the side of a hypercube is kept fixed while the number of dimensions is increased (on the contrary, expander graphs such as the RRG represent the infinite-dimensional limit taken \emph{after} the infinite-volume limit). These particular features endow the QREM with a distinguished phase diagram: at the center of the spectrum ($E=0$) the dynamics never localizes, while away from it ($E \neq 0$) a mobility edge is present. 

Here we study in detail how the unconventional mobility edge influences the RG flow. We find that, at $E=0$, the RG beta function clearly signals the absence of a transition, and the flow towards ergodicity is similar to the one observed in fully-connected models. Furthermore, the disorder term can be rescaled so that a localization transition described by a 2PS appears, i.e.\ similar to the Anderson model on expander graphs and the XXZ chain in a random field. At $E \neq 0$ the localization transition occurs without any disorder rescaling and, once again, is described by a 2PS flow. 
 
Our results are relevant for the understanding of both single-particle and many-body localization. In particular, they show what the RG flow looks like in the \emph{absence} of a localization transition: the situation appears rather different from the one observed in the random-field XXZ chain~\cite{Niedda2024}, where it was also claimed that localization is absent by some works~\cite{Suntajs2020Quantum,Sels2021Dynamical}.

\section{Model}
\begin{figure}
    \centering
    \includegraphics[width=0.8\linewidth]{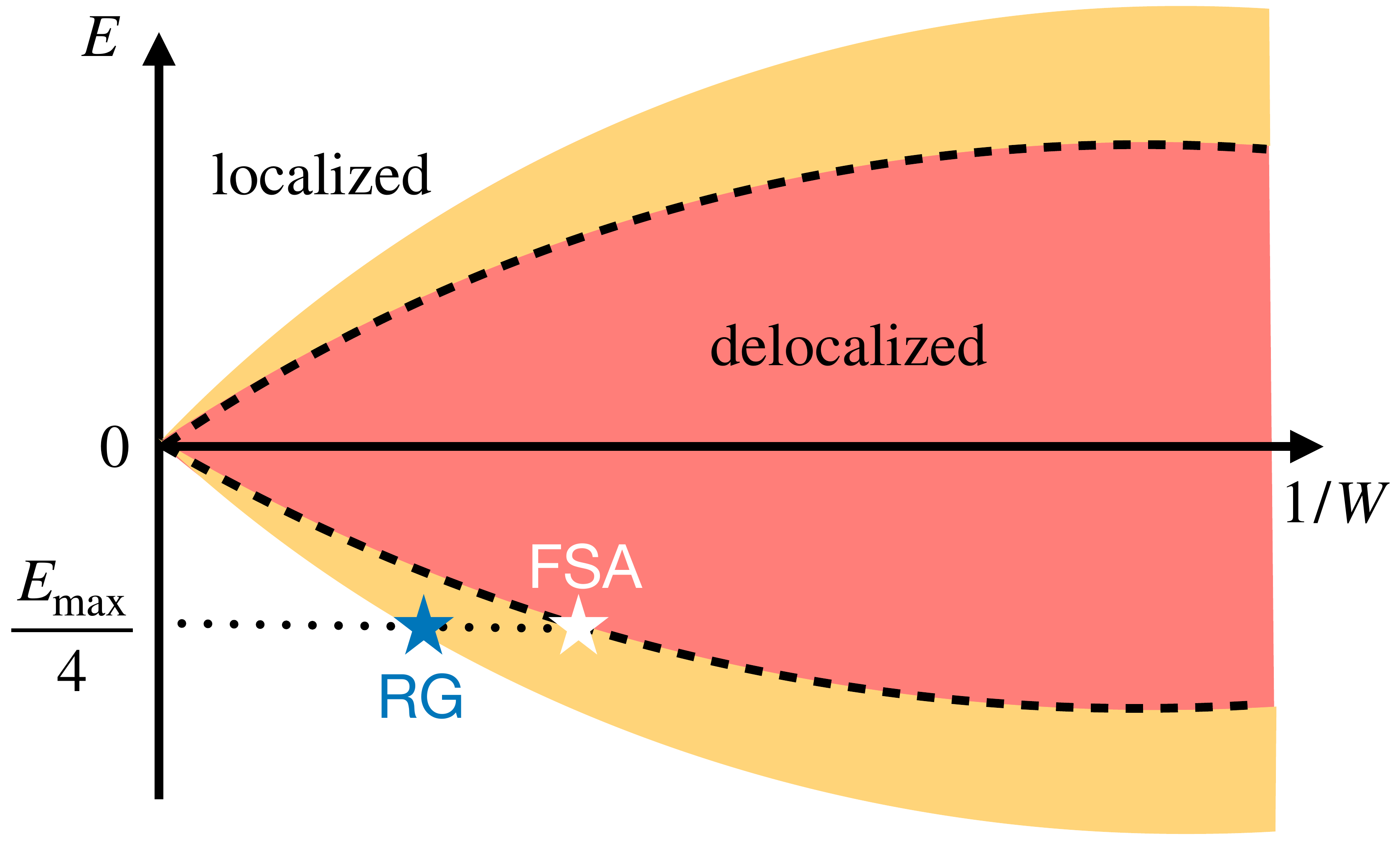}
    \caption{Cartoon of the phase diagram of the quantum random energy model in the $E,1/W$ plane. The red-shaded area indicates the delocalized region, the yellow-shaded area indicates the region that is localized in perturbation theory, but is destabilized by non-perturbative effects, and the white area is the localized phase. The dashed line corresponds to the mobility edge obtained via the forward scattering approximation (FSA). The white and blue marks are respectively the transition point predicted by FSA at $E=E_\mathrm{max}/4$ and the one that we extrapolate via renormalization group techniques. As expected, the position of the mobility edge found via exact diagonalization is at a greater value of disorder than the one given by the FSA. }
    \label{fig:PhaseDiagram}
\end{figure}

The QREM is defined by the Hamiltonian~\cite{Laumann2014MBMobility}
\begin{equation} \label{eq:Hamilt}
    H = \Gamma \sum_{i=1}^L \sigma_i^x + \mathrm{diag}\{ \epsilon_1, \dots, \epsilon_{2^L} \},
\end{equation}
where $\epsilon_j$, $j = 1, \dots, 2^L$, are i.i.d.\ gaussian random variables with average zero and standard deviation $\sqrt{L/2}$, and $\Gamma$ is the transverse field strength. In the following, we will interchangeably use also the effective disorder strength $W \equiv 1/\Gamma$. The scaling of the disorder with $\sqrt{L}$ is necessary for the Hamiltonian~\eqref{eq:Hamilt} to be extensive in $L$. 

The QREM is the quantum version of the random energy model, introduced by Derrida as an exactly solvable model arising from the limit $p \rightarrow \infty$ of $p$-spin glasses~\cite{Derrida1980REM,*Derrida1981REM}. Its classical version ($\Gamma = 0$) is typically considered as the simplest statistical mechanics model displaying a glassy phase below the critical temperature $T_\mathrm{glass} = 1/2 \sqrt{\log 2}$~\cite{GrossMezard1984}. Reintroducing quantum fluctuations, the equilibrium phase diagram of the model displays also a quantum paramagnetic phase, on top of the classical paramagnetic and glassy phases~\cite{Goldschmidt_1990,Manai2020Phase,*Manai2023Spectral}.

From a dynamical viewpoint, the system can be thought of as a particle hopping on a hypercubic lattice, in presence of a random on-site potential that is size-dependent: the model is thus strictly equivalent to the Anderson model on said geometry. Thanks to this mapping, the QREM can be also considered as a ``mean-field'' version of disordered spin chains, such as the XXZ chain in a random field: these systems can be represented as Anderson models on a Fock-space graph with correlated disorder. In the QREM, the only difference is that the disorder becomes artificially uncorrelated in the computational basis, while retaining the scaling $\sim L^{1/2}$ expected from the sum of $L$ independent disordered fields in real space.

The QREM hosts a many-body localized phase for energy densities $|E/L| > \Gamma$, as demonstrated both via perturbation theory and numerically~\cite{Laumann2014MBMobility,Baldwin2016ManyBody,Biroli2021Out,Scoquart2024Role}, see Fig.~\ref{fig:PhaseDiagram} for a sketch. This means that the critical temperature of the MBL transition is $T_{\mathrm{MBL}}=(2\Gamma)^{-1}$, much larger than the equilibrium glass transition $T_\mathrm{glass}$ for sufficiently small $\Gamma$. At the center of the spectrum, i.e.\ at zero energy density, the QREM remains ergodic regardless of the value of $\Gamma$.

The best estimate for the mobility edge $E_c(\Gamma)$ was found using a perturbative expansion of the resolvent, to estimate the probability of having a resonance at a given distance~\cite{Laumann2014MBMobility,Baldwin2016ManyBody}. Within this approach, commonly called forward scattering approximation (FSA)~\cite{pietracaprina2016forward}, the error committed is to neglect repeating paths connecting two sites, thus the approximation becomes accurate in absence of loops (e.g.\ on the Bethe lattice~\cite{abou1973selfconsistent}), and is well-controlled if there are no short-scale loops (e.g.\ on RRGs). In the case of the QREM, however, small loops are naturally present because of the hypercubic geometry, and therefore the FSA can only give a rough estimate of the critical point.

\section{Spectral observables} 

In order to probe the ergodic or localized behavior of the model, we use two different figures of merit, both calculated through exact diagonalization of the Hamiltonian in Eq.~\eqref{eq:Hamilt}~\cite{Pietracaprina2018Shift}. 

The first is the spectral gap ratio, or $r$-parameter~\cite{oganesyan2007localization}, $r = \mathbb{E} [\min(\Delta E_n,\Delta E_{n+1}) / \max(\Delta E_n,\Delta E_{n+1}) ]$, where $\Delta E_n = E_{n+1}-E_{n}$ is the $n$-th energy gap in a narrow energy window and $\mathbb{E}$ denotes the average both over the window and the disorder. Notice that the use of such a window in the ergodic regime would define the microcanonical ensemble. The value $r_{\mathrm{P}} = 2 \ln 2 - 1 \simeq 0.386$ is the Poisson value for integrable systems, while $r_{\mathrm{WD}} \simeq 0.5307$ is the Wigner-Dyson value for the Gaussian Orthogonal Ensemble (GOE), corresponding to chaotic systems with time-reversal symmetry~\cite{Atas2013Distribution}. For convenience, we rescale the average gap ratio defining $\phi = (r - r_{\mathrm{P}})/(r_{\mathrm{WD}}- r_{\mathrm{P}})$: $\phi = 0$ corresponds to a non-ergodic behavior, while $\phi=1$ to ergodicity. 

The second observable which will be used as an order parameter for the localization transition is the eigenstate fractal dimension $D$, defined as $D(L)=\rmd S(L) / \rmd \ln N$, where $N=2^L$ is the Hilbert space volume.
$S(L)$ is the average eigenfunction Shannon entropy at size $L$, i.e.\
$S(L) = - \mathbb{E} \big[ \sum_j |\braket{j}{\psi_n}|^2 \ln |\braket{j}{\psi_n}|^2 \big]$, where $j$ labels the computational basis states and $\ket{\psi_n}$ are the eigenstates within an narrow energy window as above. In the thermodynamic limit, the fractal dimension is $D=1$ for eigenfunctions extended over the whole Hilbert space, and $D=0$ for localized eigenfunctions.

\section{Scaling theory of localization}

The approach to the thermodynamic limit in the RG framework is described by the beta function, which for a generic observable $A$ can be defined as
\begin{equation}
    \label{eq:1p_scaling}
    \beta \equiv \frac{\rmd \ln A }{\rmd \ln N}.
\end{equation}
Above, the RG parameter (cutoff) has been traded with the system size dimension~\cite{Cardy1988,Cardy_1996}. This object can be reconstructed from the numerical data, in our case by replacing $A$ with the rescaled $r$-parameter $\phi$ or the fractal dimension $D$. 

In general, the beta function depends on all the relevant and irrelevant observables present in the theory. However, once the scaling regime is reached, the dependence on the irrelevant operators disappears, leaving only relevant observables to describe the proximity of the fixed points of the RG flow. If the critical regime is described by just a single relevant observable, i.e.\ there is one-parameter scaling (1PS), $\beta = \beta(A)$ holds. This happens, for instance, in the Anderson model in finite-dimensional Euclidean space~\cite{abrahams1979scaling,altshuler2024renormalization}, where the localized point, the ergodic point and the transition point are all isolated fixed points. In the case in which the critical point collides with a line of fixed points, more parameters are needed to address the critical behavior: for instance, in the BKT~\cite{Berezinskii1971Destruction,*Berezinskii1972Destruction,Kosterlitz1973Ordering} transition or in the Anderson model in infinite dimensions~\cite{vanoni2023renormalization} (and probably in interacting systems as well~\cite{Niedda2024}) there are one relevant and one marginal operator. As a consequence, two equations are needed to describe the RG flow in the neighborhood of some critical points. In this case, the general equations describing two-parameter scaling (2PS) can be written using $\beta$ itself as a local coordinate:
\begin{equation}  \label{eq:2PS}
    \begin{cases}
        \dot{\alpha} = \beta, \\
        \dot{\beta} = \gamma(\alpha, \beta),
    \end{cases}
\end{equation}
where $\gamma$ is some function, and we put $t \equiv \ln N$ and $\alpha \equiv \ln A$. 

We now specialize the discussion to the {\it localized} critical point, which we assume to be characterized by the value $A=0$, thus $\alpha\to-\infty$. By the request of being a fixed point, $\beta=0$ and, assuming the function $\gamma(\alpha,\beta)$ to be regular, one can expand $\gamma(\alpha,\beta)=\gamma(\alpha,0)+\beta \gamma_1(\alpha)+O(\beta^2).$ This regularity assumption is supported by numerical data in the known cases~\cite{Niedda2024,vanoni2023renormalization}. The approximation of replacing $\gamma(\alpha, \beta) \rightarrow \gamma(\alpha,0)$ endows the system of equations with a symplectic structure and allows for the mapping of the RG flow onto the orbits of a one-dimensional Hamiltonian particle of coordinate $\alpha$, subjected to an external potential $V(\alpha) = - \int_{\alpha_0}^\alpha d\alpha' \gamma(\alpha')$. In this formulation, the absence or presence of a transition correspond to a confining or non-confining potential, respectively. The orbits can be parametrized by the conserved energy $\mathcal{E}=\dot{\alpha}^2/2 + V(\alpha)$ of the effective particle.

Confining potentials do not allow orbits to flow to $\alpha\to -\infty$ (i.e.\ $A=0$) for any value of the initial conditions. The particle will always be repelled by the localized line of fixed points and flow towards positive $\alpha$: this situation corresponds to systems which possess only a stable {\it ergodic phase}. On the other hand, non-confining potentials allow for the orbits to flow to the localized value $\alpha\to-\infty$ (i.e.\ $A=0$) when the effective energy $\mathcal{E}$ exceeds some critical value, necessary to escape the potential wall. For these orbits, by assuming an exponential decay of the observable $A \propto e^{-L/\xi_{\rm loc}}$, the second line in Eq.~\eqref{eq:2PS} describes the renormalization of the localization length $\xi_{\rm loc}$ and the attachment of the RG flow to a line of localized fixed points on the semi-axis $(A=0,\, \beta=-1/\xi_{loc})$. The localization length can be related to the residual momentum of the particle at $\alpha\to-\infty$.

\section{Numerical results}

\begin{figure*}
    \centering
    \includegraphics[height=0.21\textwidth]{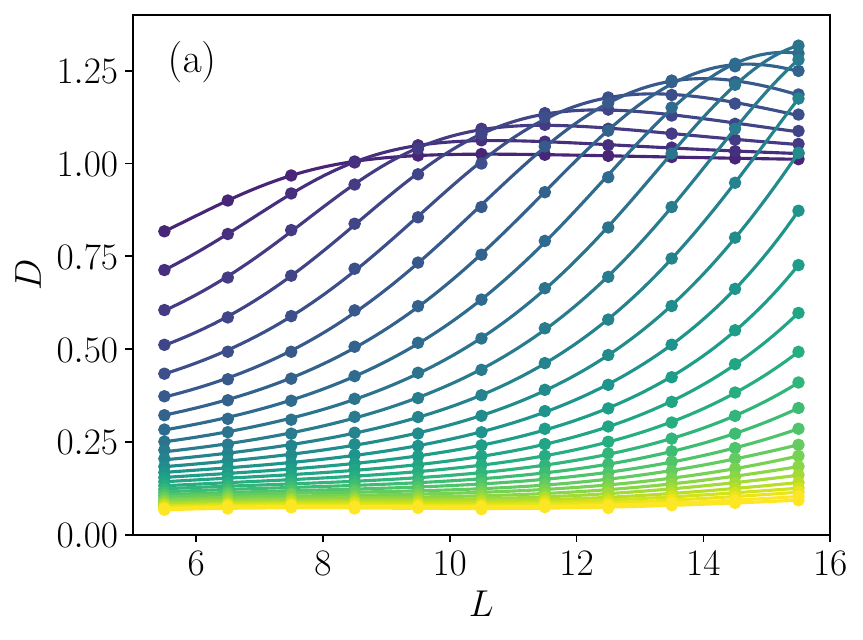}
    \includegraphics[height=0.21\textwidth]{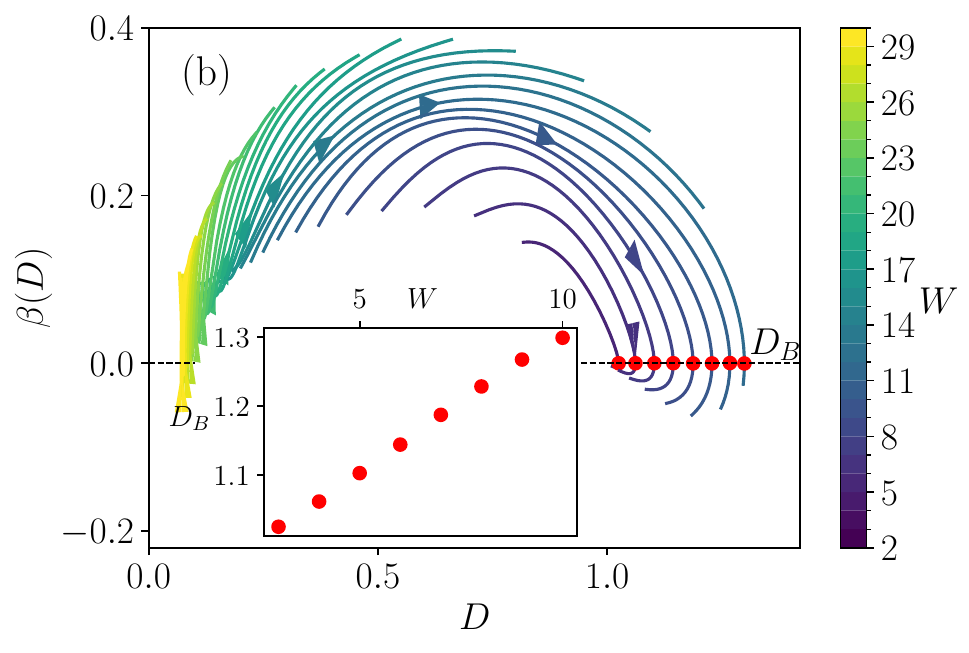}
    \includegraphics[height=0.20\textwidth]{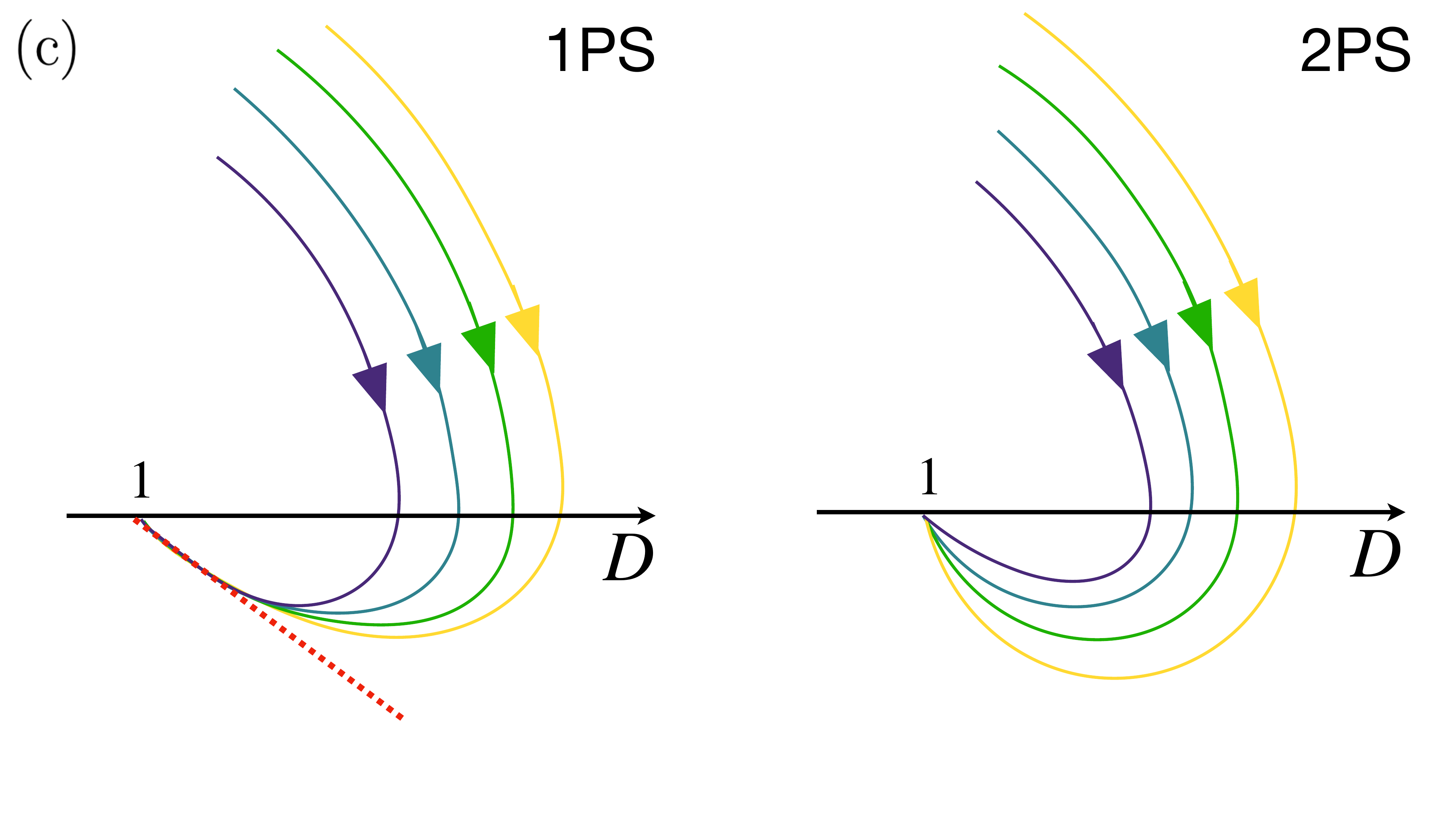}
    \caption{(a) System-size dependence of the fractal dimension $D$ at the center of the spectrum (i.e.\ at infinite temperature). Dots are the numerical data, obtained by taking the finite-difference derivative of the participation entropy, and solid lines are the interpolation of the data, used for computing the beta function. For any $W$, the curves will eventually reach $D=1$ for sufficiently large system size, as all curves flow towards increasing values of $D$ as the system size grows. At intermediate $L$, the fractal dimension becomes $D>1$, signaling that the eigenfunction support set grows faster than the system size.
    (b) Beta function at $E=0$, with direction of the RG flow indicated by the arrows. The red dots in the main plot are the zeros of the beta function at $D>1$, and we denote with $D_B$ the corresponding value of $D$. In the inset, we report the dependence of $D_B$ on $W$: a linear behavior is observed, suggesting the persistence of this effect for large sizes. (c) Pictorial representation of the two possible scenarios for the RG flow at the ergodic fixed point, one-parameter (1PS) or two-parameter (2PS) scaling. The red dashed line is a sketch for the 1PS asymptotic curve.
    }
    \label{fig:D_QREM}
\end{figure*}

Numerical data qualitatively agree with the picture coming from perturbation theory~\cite{Laumann2014MBMobility,Baldwin2016ManyBody}. The fractal dimension at the center of the spectrum, $E=0$, is reported as a function of system size in Fig.~\ref{fig:D_QREM}(a), and the corresponding beta function in Fig.~\ref{fig:D_QREM}(b): all curves flow to the ergodic fixed point for any value of $W$, as long as one keeps $W=O(1)$. The same conclusion can be reached from the $r$-parameter data, that we report in the Supplement~\cite{SupplMat}. Notice that the 2PS cannot be inferred from the flow in the $(r,D)$ plane, as the two parameters are a function of each other~\cite{vanoni2023renormalization}. We anticipate that one can observe a transition for $E=0$, provided one rescales the disorder as already pointed out in Refs.~\cite{Laumann2014MBMobility,Baldwin2016ManyBody}. We will explore this transition in the next section.

As one can notice, in the ergodic regime the fractal dimension reaches the ergodic fixed point $D=1$ from \emph{above}, signaling that the support set of the wave functions is growing faster than the Hilbert space dimension. While in the thermodynamic limit the fractal dimension must be smaller than (or at most equal to) one, the appearance of $D(L)>1$ at finite size was observed also in the Anderson model in $d=2$~\cite{altshuler2024renormalization}, in Josephson junction arrays~\cite{Pino2017Multifractal} and, more interestingly, in Gaussian and log-normal Rosenzweig-Porter models~\footnote{This behavior can be deduced from the data reported in Ref.~\cite{kutlin2024investigating}, even if the fact is not explicitly commented.}. We conjecture that this behavior is common to models whose connectivity grows with the system size and whose disorder is uncorrelated in Hilbert space.

Because of the overshooting described above, the beta function of the fractal dimension approaches the ergodic fixed point in a peculiar way. The zeros of $\beta$ at $D > 1$, which correspond to the maxima of $D$ as a function of $L$ (see Fig.~\ref{fig:D_QREM}(a)), increase with $W$, as shown in the inset of Fig.~\ref{fig:D_QREM}(b): it is reasonable to assume that this behavior may persist even at very large disorder and system sizes. Although the fractal dimension eventually reaches $D=1$ in the thermodynamic limit, the approach to the ergodic fixed point may follow either 1PS or 2PS, depending on whether there is a residual dependence of the slope on $W$, see Fig.~\ref{fig:D_QREM}(c) for a sketch of these two cases. The numerical data available do not allow us to conclude which scenario is correct. However, as will be discussed in the following, the 1PS to the ergodic fixed point is recovered both if the disorder rescaling is introduced, and away from the center of the spectrum. This suggests that the 1PS scenario is most likely the correct one for the RG flow at the ergodic fixed point.

\section{Rescaling the disorder} 
\label{sec:rescaling}

\begin{figure*}
    \centering
    \includegraphics[height=0.223\textwidth]{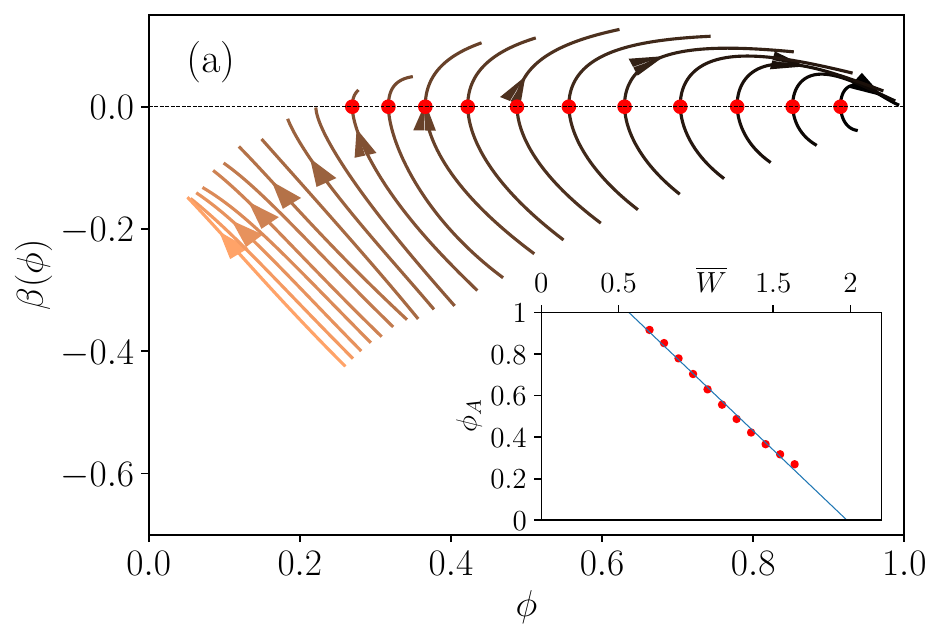}
    \includegraphics[height=0.227\textwidth]{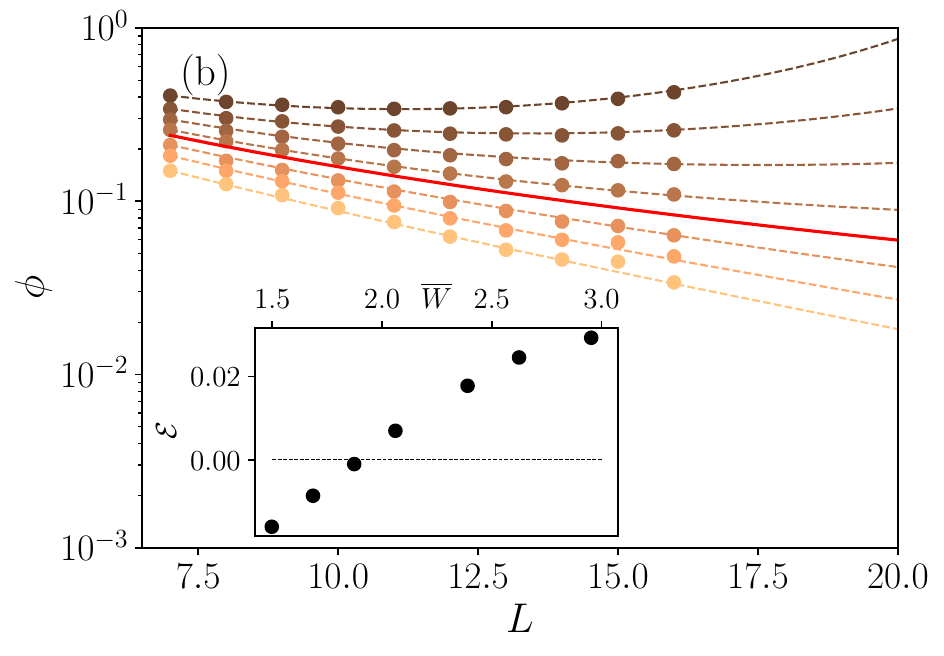}
    \includegraphics[height=0.223\textwidth]{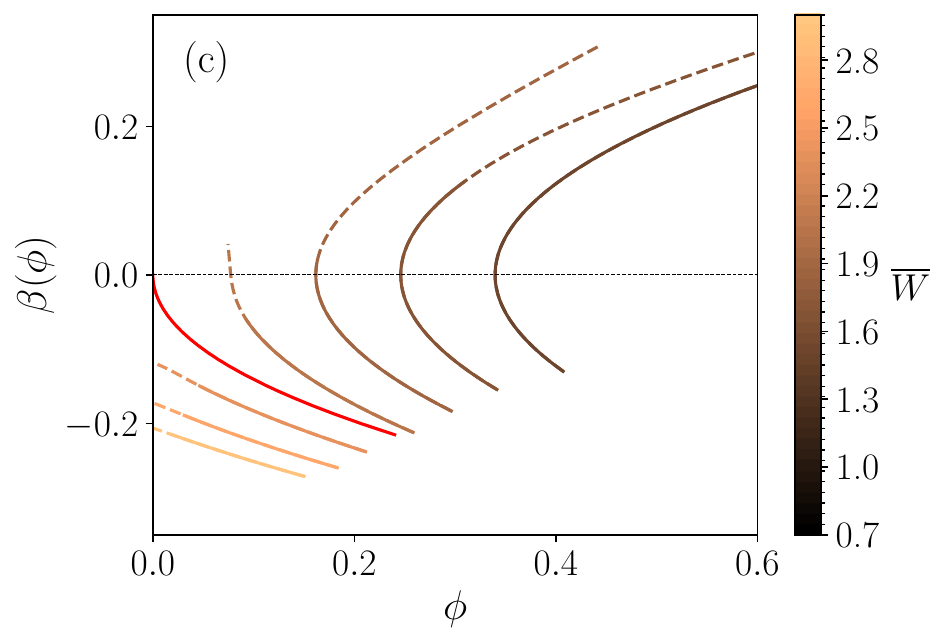}
    \caption{(a) Beta function of the rescaled $r$-parameter $\phi$, parametrized by the rescaled disorder strength $\overline{W}= W/\sqrt{L}\log L$. The flow is consistent with the presence of a line of fixed points on the negative $\beta$ axis, see also panel (c). In the inset, the turning points of the RG dynamics are shown as a function of $\overline{W}$: a linear fit extrapolates to $\phi_c=0$ at $\overline{W}_c\simeq 1.95$. (b) Fits of the numerical data obtained from the orbits of the dynamical system with potential $V(\alpha)=-(c/n)e^{n\alpha}$ and parameters $c \simeq 0.05$ and $n \simeq 1$. In the inset, the energies labeling the orbits are plotted as a function of  $\overline{W}$. (c) Phase-space plot of the dynamical system orbits: lines are continuous up to the RG time at which data are available, and become dashed along the extrapolations to higher sizes.  
    }
    \label{fig:beta_rescaled_W}
\end{figure*} 

While the QREM is always ergodic at the center of the spectrum, a localization transition can be induced by rescaling the disorder to $\overline{W}= W / \sqrt{L} \log L$. The additional logarithmic factor arises from the careful treatment of the perturbative expansion~\cite{Laumann2014MBMobility,Baldwin2016ManyBody}, following the seminal example of the Bethe lattice~\cite{abou1973selfconsistent}. With this rescaling, the diagonal term becomes $O(L)$ from the single-particle perspective, and compensates for the $O(L)$ connectivity of the Fock space graph: it is a form of Kac rescaling~\cite{Kac1956Foundations,Campa2014Physics}. However, from the perspective of spins, the extensivity of the energy is no longer maintained. This non-extensivity of energy is not a cause for concern, as localization is fundamentally a dynamical phase transition. On the other hand, the scaling required to ensure thermodynamic extensivity prevents the observation of the localization transition at the center of the spectrum.

We plot the beta function of the rescaled $r$-parameter, parametrized by the rescaled disorder $\overline{W}$, in Fig.~\ref{fig:beta_rescaled_W}. While bigger system sizes are needed to draw definitive conclusions on the nature of the transition, it appears plausible that it is described by a 2PS theory, with a line of fixed points on the negative $\beta$ axis. The situation appears similar to the Anderson transition on expander graphs~\cite{vanoni2023renormalization}, and potentially to the random-field XXZ spin chain~\cite{Niedda2024}. The approach to the ergodic fixed point follows instead a 1PS curve, as the beta function curves collapse on one another.

The turning points $\phi_A$ of the RG dynamics, i.e.\ the minima of $\phi$, decrease linearly to $\phi_c=0$ at the finite value $\overline{W}_c \simeq 1.95$, when plotted as a function of $\overline{W}$ (see inset of Fig.~\ref{fig:beta_rescaled_W}(a)). 
Remarkably, our result is in very good agreement with the analytical prediction obtained in Ref.~\cite{Scoquart2024Role}. In fact, according to their derivation, and after an appropriate rescaling to cast the result of Ref.~\cite{Scoquart2024Role} in our same convention, the theoretical prediction of the rescaled critical disorder is $\overline{W}_c \simeq 4/\sqrt{\pi} \simeq 2.25$, to be compared with our numerical prediction $\overline{W}_c \simeq 1.95$.
In addition, it is interesting to notice that the analytical prediction presented in Ref.~\cite{Scoquart2024Role} is obtained by considering the QREM to be equivalent to an RRG of coordination number $L$ and volume $2^L$. Given that the universality class of the Anderson model on the RRG should not depend on the coordination of the graph~\cite{sierant2023universality,vanoni2023renormalization}, the above observation motivates the observation that the QREM at $E=0$ belongs to the same universality class as the RRG. Let us also mention that, with our definitions of the disorder strength, the critical disorder on a Bethe lattice would be $\overline{W}_c^{\mathrm{BL}} = 2 / \sqrt{\pi} \simeq 1.12$~\cite{parisi2019anderson}. It is possible that the value in the QREM turns out to be larger because of the presence of loops in the Fock-space graph, which favor resonances. Here, loops are particularly important because the number of dimensions of the Fock space is increased while keeping the side of the hypercubic lattice fixed to 1. The Bethe lattice, instead, is reached by keeping the number of spins fixed and increasing the spin representation: this is equivalent to considering the Anderson model in finite $d$-dimensional space~\cite{altshuler2024renormalization} and taking the limit $d \rightarrow \infty$. A possible way to interpolate the two situations would be to consider, instead of spins-1/2, an arbitrary spin $S$ which generates an $L$-dimensional hypercube with $(2S+1)^L$ vertices of size $2S+1$ in each direction. We conjecture that the critical value $\overline{W}_c(S)\to \overline{W}_c^{\mathrm{BL}}$ as $S\to\infty$.

It is interesting to study the RG dynamics around the transition and in the localized phase via the effective dynamical system introduced above. In order to fit the data, we use the orbits of a Hamiltonian system with potential $V(\alpha)=-(c/n)e^{n\alpha}$, where $\alpha = \ln \phi$; see Fig.~\ref{fig:beta_rescaled_W}(b,c). Each curve is obtained with a different choice of the initial conditions of the dynamics, which can be used to compute the effective energies $\mathcal{E}=\beta^2/2 - (c/n) e^{n\alpha}$. The best fits yield $c \simeq 0.05$ and $n \simeq 1$; the exponent $n$, characterizing the scaling on the critical line $\phi(L)\sim L^{-2/n}$, has remarkably the same value as in the Anderson model on RRG~\cite{vanoni2023renormalization}, and it is also consistent with what observed in the XXZ spin chain~\cite{Niedda2024}. The energies of each trajectory are in one-to-one correspondence with the disorder values $\overline{W}$. The value of $\overline{W}$ at which the energies cross the horizontal axis $\mathcal{E}=0$ corresponds to the critical point: a linear fit yields a critical disorder $\overline{W}_c \simeq 1.9$, which is consistent with the value obtained from the turning points $\phi_A$. The critical line can be obtained by integrating the equations of motion for $\mathcal{E}=0$ and is marked with red in the figure. A line of fixed points becomes apparent for the larger values of $\overline{W}$, leading to a 2PS behavior with one relevant and one marginal direction

\section{On the mobility edge}

\begin{figure*}
    \centering
    \includegraphics[width=0.35\linewidth]{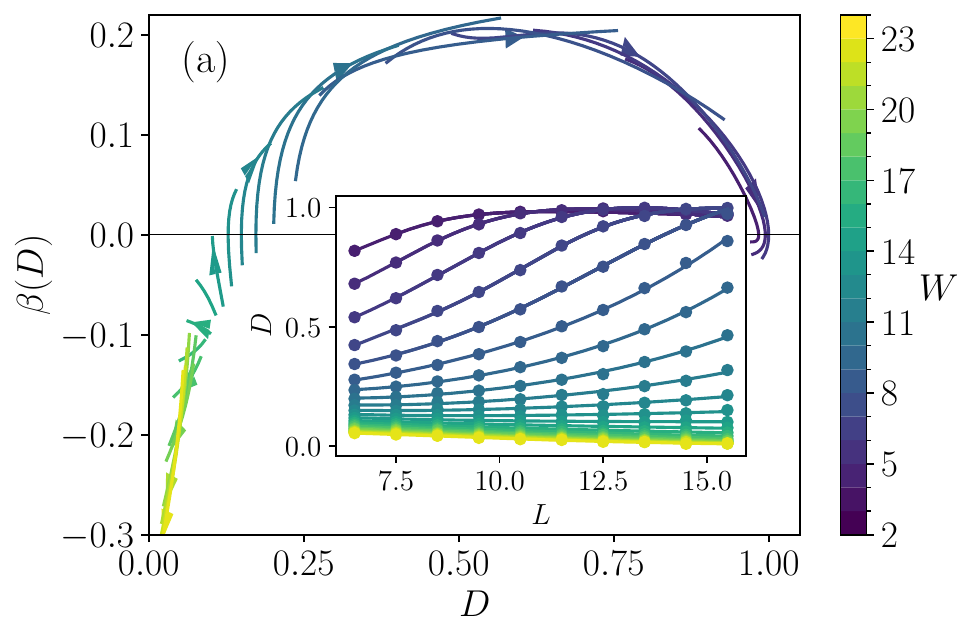}
    \includegraphics[width=0.3\linewidth]{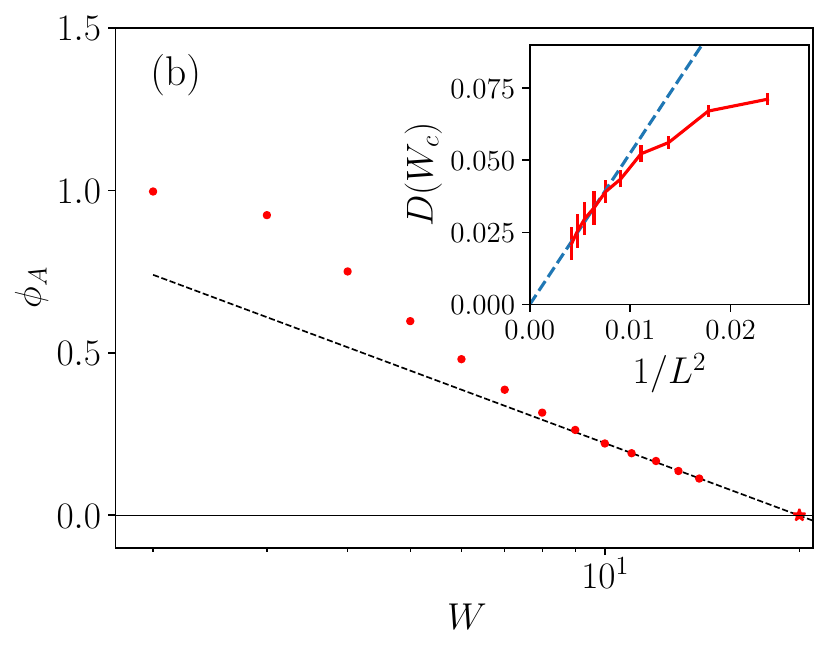}
    \caption{(a) Beta function (main) and the corresponding fractal dimension (inset) away from the center of the spectrum, at $E=E_{\mathrm{max}}/4$ with $E_{\mathrm{max}}$ largest energy eigenvalue. The approach to $D=1$ suffers from less severe finite-size effects, and $D$ only slightly exceeds 1. Moreover, at large $W$ the curves $D(L)$ approach $D=0$, signaling the presence of the localized phase.
    (b) Determination of the critical value $W_c$ at energy $E=E_{\mathrm{max}}/4$: $\phi_A$ is such that $\beta(\phi_A) = 0$, and it extrapolates to $\phi_A = 0$ for $W_c \simeq 20$. (Inset) At $W=W_c$, quantities as the fractal dimension $D$ vanish as $1/L^2$, a behavior typical of expander graphs~\cite{sierant2023universality,vanoni2023renormalization}.
    }
    \label{fig:D_QREM_Wc_finiteT}
\end{figure*}

Away from the center of the spectrum, numerical data confirm the presence of a mobility edge, see Fig.~\ref{fig:D_QREM_Wc_finiteT}(a). We argue that the beta function behaves, in the large systems size limit, similarly to the one of the Anderson model on RRGs, for the following reason.~At finite energy density, the density of states is smaller than at infinite temperature, and the subset of resonant sites (for $\Gamma=0$) is exponentially smaller than the complete Hilbert space. In general, the resonant sites will be distributed randomly on the hypercube, in a fashion similar to a random graph. Turning on the hopping, one expects therefore the localization transition universality class to be the same of the RRG~\cite{sierant2023universality}. Finite-size effects which are regulated by irrelevant operators, however, can in general be different, as one can notice from comparing the beta function in the localized region to the RRG result. More about this can be found in the Supplement~\cite{SupplMat}.

In the main panel of Fig.~\ref{fig:D_QREM_Wc_finiteT}(b), we plot the turning points of the rescaled $r$-parameter as a function of the disorder. A linear extrapolation to the critical value $\phi=0$ yields the critical disorder $W_c \simeq 20$. The FSA presented in Ref.~\cite{Baldwin2016ManyBody} gives $\Gamma_c \simeq 0.283$ for energy densities $\epsilon = \sqrt{\log(2)}/4$ \footnote{Notice that $\epsilon_0 = \pm \sqrt{\log(2)}$ is the ground-state energy of the classical REM, that approximates well the edge of the spectrum $E_{\mathrm{max}}$. The value of $\Gamma_c$ has been obtained using Eq. (35) of Ref.~\cite{Baldwin2016ManyBody}.}, corresponding to $W_c^{\mathrm{FSA}} \simeq 3.5$.
In the inset, the scaling of the fractal dimension on the critical line is shown: data are compatible with $D(L,W_c)\sim L^{-2/n}$, $n\simeq 1$, which corresponds to what we find integrating the equation of motion with the potential $V(\delta) = - (c/n) e^{n \delta}$, with $\delta = \ln D$. This is the same potential characterizing the scaling of the rescaled $r$-parameter on the critical line at the spectrum center, after the disorder rescaling. Therefore, the universality class of the localization transition at finite energy is identical to that at the center of the spectrum, exemplified paradigmatically by the Anderson model on RRGs.

Finally, we notice that the rescaling of disorder considered in Sec.~\ref{sec:rescaling}, which makes the localization transition appear at the center of the spectrum, changes as well the scenario at finite energy density: the system becomes always localized, as a finite value of $W$ in the thermodynamic limit corresponds to $\overline{W} = 0$.

\section{Discussion} 

The scaling theory of the localization transition in the QREM has been studied in the present work by reconstructing numerically the beta function of spectral observables. In the ``natural'' scaling of the disorder variance, which makes the model a mean-field version of MBL models, our analysis confirms the absence of transition at the center of the spectrum and opens two possible scenarios for the approach to the ergodic fixed point: either a one- or a two-parameter scaling. On the other hand, at finite energy density a localization transition is described by a two-parameter scaling theory similarly to the Anderson model on expander graphs. It was noted in Ref. \cite{FAORO2019167916} that a different microcanonical phase diagram was proposed, suggesting the existence of a non-ergodic extended phase at finite energy density. We emphasize that our findings are not necessarily in conflict with this conjecture. However, a definitive confirmation of such a phase would require access to larger system sizes, which are essential for a more reliable extrapolation of the $\beta$ function within the ergodic regime.

A transition to a localized phase emerges also at the center of the spectrum if one suitably rescales the variance of the disorder: the transition appears to be described by a two-parameter scaling theory even in this case, with the same critical exponent of the Anderson model on the RRG and possibly of the XXZ spin chain. Our results show that the universality class of the localization transition in the QREM is the same of the Anderson model on any random graph and, remarkably, does not change by rescaling the disorder. This is a consequence of the fact that the RG is blind to the scaling of the microscopic variables.

\acknowledgments

We acknowledge discussion with C.\ L.\ Baldwin, I.\ V.\ Gornyi, D.\ A.\ Huse, C.\ R.\ Laumann, A.\ D.\ Mirlin and T.\ Scoquart, and we thank B.\ L.\ Altshuler and V.\ E.\ Kravtsov for collaborations on related topics. We are also grateful to one of the Referees for pointing out the agreement between our results and those of Ref.~\cite{Scoquart2024Role}. Numerical simulations were performed using the libraries SciPy~\cite{SciPy}, PETSc~\cite{PETSc} and SLEPc~\cite{SLEPc} on the MPIPKS HPC cluster. This work was partly funded by the European Union -- NextGenerationEU under the project NRRP “National Centre for HPC, Big Data and Quantum Computing (HPC)'' CN00000013 (CUP D43C22001240001) [MUR Decree n.\ 341--15/03/2022] -- Cascade Call launched by SPOKE 10 POLIMI: ``CQEB'' project. The work of JN was funded by the European Union--NextGenerationEU under the project NRRP Project ``National Quantum Science and Technology Institute" — NQSTI, Award Number: PE00000023, Concession Decree No.~1564 of 11.10.2022 adopted by the Italian Ministry of Research, CUP J97G22000390007.


\bibliography{references}

\begin{thebibliography}{82}%
\makeatletter
\providecommand \@ifxundefined [1]{%
 \@ifx{#1\undefined}
}%
\providecommand \@ifnum [1]{%
 \ifnum #1\expandafter \@firstoftwo
 \else \expandafter \@secondoftwo
 \fi
}%
\providecommand \@ifx [1]{%
 \ifx #1\expandafter \@firstoftwo
 \else \expandafter \@secondoftwo
 \fi
}%
\providecommand \natexlab [1]{#1}%
\providecommand \enquote  [1]{``#1''}%
\providecommand \bibnamefont  [1]{#1}%
\providecommand \bibfnamefont [1]{#1}%
\providecommand \citenamefont [1]{#1}%
\providecommand \href@noop [0]{\@secondoftwo}%
\providecommand \href [0]{\begingroup \@sanitize@url \@href}%
\providecommand \@href[1]{\@@startlink{#1}\@@href}%
\providecommand \@@href[1]{\endgroup#1\@@endlink}%
\providecommand \@sanitize@url [0]{\catcode `\\12\catcode `\$12\catcode
  `\&12\catcode `\#12\catcode `\^12\catcode `\_12\catcode `\%12\relax}%
\providecommand \@@startlink[1]{}%
\providecommand \@@endlink[0]{}%
\providecommand \url  [0]{\begingroup\@sanitize@url \@url }%
\providecommand \@url [1]{\endgroup\@href {#1}{\urlprefix }}%
\providecommand \urlprefix  [0]{URL }%
\providecommand \Eprint [0]{\href }%
\providecommand \doibase [0]{https://doi.org/}%
\providecommand \selectlanguage [0]{\@gobble}%
\providecommand \bibinfo  [0]{\@secondoftwo}%
\providecommand \bibfield  [0]{\@secondoftwo}%
\providecommand \translation [1]{[#1]}%
\providecommand \BibitemOpen [0]{}%
\providecommand \bibitemStop [0]{}%
\providecommand \bibitemNoStop [0]{.\EOS\space}%
\providecommand \EOS [0]{\spacefactor3000\relax}%
\providecommand \BibitemShut  [1]{\csname bibitem#1\endcsname}%
\let\auto@bib@innerbib\@empty
\bibitem [{\citenamefont {Anderson}(1958)}]{Anderson1958absence}%
  \BibitemOpen
  \bibfield  {author} {\bibinfo {author} {\bibfnamefont {P.~W.}\ \bibnamefont
  {Anderson}},\ }\href {https://doi.org/10.1103/PhysRev.109.1492} {\bibfield
  {journal} {\bibinfo  {journal} {Phys. Rev.}\ }\textbf {\bibinfo {volume}
  {109}},\ \bibinfo {pages} {1492} (\bibinfo {year} {1958})}\BibitemShut
  {NoStop}%
\bibitem [{\citenamefont {Evers}\ and\ \citenamefont
  {Mirlin}(2008)}]{evers2008anderson}%
  \BibitemOpen
  \bibfield  {author} {\bibinfo {author} {\bibfnamefont {F.}~\bibnamefont
  {Evers}}\ and\ \bibinfo {author} {\bibfnamefont {A.~D.}\ \bibnamefont
  {Mirlin}},\ }\href {https://doi.org/10.1103/RevModPhys.80.1355} {\bibfield
  {journal} {\bibinfo  {journal} {Rev. Mod. Phys.}\ }\textbf {\bibinfo {volume}
  {80}},\ \bibinfo {pages} {1355} (\bibinfo {year} {2008})}\BibitemShut
  {NoStop}%
\bibitem [{\citenamefont {Slevin}\ and\ \citenamefont
  {Ohtsuki}(1999)}]{Slevin-PRL99}%
  \BibitemOpen
  \bibfield  {author} {\bibinfo {author} {\bibfnamefont {K.}~\bibnamefont
  {Slevin}}\ and\ \bibinfo {author} {\bibfnamefont {T.}~\bibnamefont
  {Ohtsuki}},\ }\href {https://doi.org/10.1103/PhysRevLett.82.382} {\bibfield
  {journal} {\bibinfo  {journal} {Phys. Rev. Lett.}\ }\textbf {\bibinfo
  {volume} {82}},\ \bibinfo {pages} {382} (\bibinfo {year} {1999})}\BibitemShut
  {NoStop}%
\bibitem [{\citenamefont {Ueoka}\ and\ \citenamefont
  {Slevin}(2014)}]{ueoka2014dimensional}%
  \BibitemOpen
  \bibfield  {author} {\bibinfo {author} {\bibfnamefont {Y.}~\bibnamefont
  {Ueoka}}\ and\ \bibinfo {author} {\bibfnamefont {K.}~\bibnamefont {Slevin}},\
  }\href {https://doi.org/10.7566/JPSJ.83.084711} {\bibfield  {journal}
  {\bibinfo  {journal} {J. Phys. Soc. Japan}\ }\textbf {\bibinfo {volume}
  {83}},\ \bibinfo {pages} {084711} (\bibinfo {year} {2014})}\BibitemShut
  {NoStop}%
\bibitem [{\citenamefont {Tarquini}\ \emph {et~al.}(2017)\citenamefont
  {Tarquini}, \citenamefont {Biroli},\ and\ \citenamefont
  {Tarzia}}]{Tarquini2017critical}%
  \BibitemOpen
  \bibfield  {author} {\bibinfo {author} {\bibfnamefont {E.}~\bibnamefont
  {Tarquini}}, \bibinfo {author} {\bibfnamefont {G.}~\bibnamefont {Biroli}},\
  and\ \bibinfo {author} {\bibfnamefont {M.}~\bibnamefont {Tarzia}},\ }\href
  {https://doi.org/10.1103/PhysRevB.95.094204} {\bibfield  {journal} {\bibinfo
  {journal} {Phys. Rev. B}\ }\textbf {\bibinfo {volume} {95}},\ \bibinfo
  {pages} {094204} (\bibinfo {year} {2017})}\BibitemShut {NoStop}%
\bibitem [{\citenamefont {Fleishman}\ and\ \citenamefont
  {Anderson}(1980)}]{fleishman1980interactions}%
  \BibitemOpen
  \bibfield  {author} {\bibinfo {author} {\bibfnamefont {L.}~\bibnamefont
  {Fleishman}}\ and\ \bibinfo {author} {\bibfnamefont {P.~W.}\ \bibnamefont
  {Anderson}},\ }\href {https://doi.org/10.1103/PhysRevB.21.2366} {\bibfield
  {journal} {\bibinfo  {journal} {Phys. Rev. B}\ }\textbf {\bibinfo {volume}
  {21}},\ \bibinfo {pages} {2366} (\bibinfo {year} {1980})}\BibitemShut
  {NoStop}%
\bibitem [{\citenamefont {Altshuler}\ \emph {et~al.}(1997)\citenamefont
  {Altshuler}, \citenamefont {Gefen}, \citenamefont {Kamenev},\ and\
  \citenamefont {Levitov}}]{altshuler1997quasiparticle}%
  \BibitemOpen
  \bibfield  {author} {\bibinfo {author} {\bibfnamefont {B.~L.}\ \bibnamefont
  {Altshuler}}, \bibinfo {author} {\bibfnamefont {Y.}~\bibnamefont {Gefen}},
  \bibinfo {author} {\bibfnamefont {A.}~\bibnamefont {Kamenev}},\ and\ \bibinfo
  {author} {\bibfnamefont {L.~S.}\ \bibnamefont {Levitov}},\ }\href
  {https://doi.org/10.1103/PhysRevLett.78.2803} {\bibfield  {journal} {\bibinfo
   {journal} {Phys. Rev. Lett.}\ }\textbf {\bibinfo {volume} {78}},\ \bibinfo
  {pages} {2803} (\bibinfo {year} {1997})}\BibitemShut {NoStop}%
\bibitem [{\citenamefont {Gornyi}\ \emph {et~al.}(2005)\citenamefont {Gornyi},
  \citenamefont {Mirlin},\ and\ \citenamefont
  {Polyakov}}]{Gornyi2005Interacting}%
  \BibitemOpen
  \bibfield  {author} {\bibinfo {author} {\bibfnamefont {I.~V.}\ \bibnamefont
  {Gornyi}}, \bibinfo {author} {\bibfnamefont {A.~D.}\ \bibnamefont {Mirlin}},\
  and\ \bibinfo {author} {\bibfnamefont {D.~G.}\ \bibnamefont {Polyakov}},\
  }\href {https://doi.org/10.1103/PhysRevLett.95.206603} {\bibfield  {journal}
  {\bibinfo  {journal} {Phys. Rev. Lett.}\ }\textbf {\bibinfo {volume} {95}},\
  \bibinfo {pages} {206603} (\bibinfo {year} {2005})}\BibitemShut {NoStop}%
\bibitem [{\citenamefont {Basko}\ \emph {et~al.}(2006)\citenamefont {Basko},
  \citenamefont {Aleiner},\ and\ \citenamefont {Altshuler}}]{Basko06}%
  \BibitemOpen
  \bibfield  {author} {\bibinfo {author} {\bibfnamefont {D.}~\bibnamefont
  {Basko}}, \bibinfo {author} {\bibfnamefont {I.}~\bibnamefont {Aleiner}},\
  and\ \bibinfo {author} {\bibfnamefont {B.}~\bibnamefont {Altshuler}},\ }\href
  {https://doi.org/10.1016/j.aop.2005.11.014} {\bibfield  {journal} {\bibinfo
  {journal} {Ann. Phys. (Amsterdam)}\ }\textbf {\bibinfo {volume} {321}},\
  \bibinfo {pages} {1126} (\bibinfo {year} {2006})}\BibitemShut {NoStop}%
\bibitem [{\citenamefont {Oganesyan}\ and\ \citenamefont
  {Huse}(2007)}]{oganesyan2007localization}%
  \BibitemOpen
  \bibfield  {author} {\bibinfo {author} {\bibfnamefont {V.}~\bibnamefont
  {Oganesyan}}\ and\ \bibinfo {author} {\bibfnamefont {D.~A.}\ \bibnamefont
  {Huse}},\ }\href {https://doi.org/10.1103/PhysRevB.75.155111} {\bibfield
  {journal} {\bibinfo  {journal} {Phys. Rev. B}\ }\textbf {\bibinfo {volume}
  {75}},\ \bibinfo {pages} {155111} (\bibinfo {year} {2007})}\BibitemShut
  {NoStop}%
\bibitem [{\citenamefont {Pal}\ and\ \citenamefont {Huse}(2010)}]{Pal10}%
  \BibitemOpen
  \bibfield  {author} {\bibinfo {author} {\bibfnamefont {A.}~\bibnamefont
  {Pal}}\ and\ \bibinfo {author} {\bibfnamefont {D.~A.}\ \bibnamefont {Huse}},\
  }\href {https://doi.org/10.1103/PhysRevB.82.174411} {\bibfield  {journal}
  {\bibinfo  {journal} {Phys. Rev. B}\ }\textbf {\bibinfo {volume} {82}},\
  \bibinfo {pages} {174411} (\bibinfo {year} {2010})}\BibitemShut {NoStop}%
\bibitem [{\citenamefont {Bardarson}\ \emph {et~al.}(2012)\citenamefont
  {Bardarson}, \citenamefont {Pollmann},\ and\ \citenamefont
  {Moore}}]{Bardarson2012Unbounded}%
  \BibitemOpen
  \bibfield  {author} {\bibinfo {author} {\bibfnamefont {J.~H.}\ \bibnamefont
  {Bardarson}}, \bibinfo {author} {\bibfnamefont {F.}~\bibnamefont
  {Pollmann}},\ and\ \bibinfo {author} {\bibfnamefont {J.~E.}\ \bibnamefont
  {Moore}},\ }\href {https://doi.org/10.1103/PhysRevLett.109.017202} {\bibfield
   {journal} {\bibinfo  {journal} {Phys. Rev. Lett.}\ }\textbf {\bibinfo
  {volume} {109}},\ \bibinfo {pages} {017202} (\bibinfo {year}
  {2012})}\BibitemShut {NoStop}%
\bibitem [{\citenamefont {Serbyn}\ \emph
  {et~al.}(2013{\natexlab{a}})\citenamefont {Serbyn}, \citenamefont
  {Papi\'{c}},\ and\ \citenamefont {Abanin}}]{Serbyn2013Universal}%
  \BibitemOpen
  \bibfield  {author} {\bibinfo {author} {\bibfnamefont {M.}~\bibnamefont
  {Serbyn}}, \bibinfo {author} {\bibfnamefont {Z.}~\bibnamefont {Papi\'{c}}},\
  and\ \bibinfo {author} {\bibfnamefont {D.~A.}\ \bibnamefont {Abanin}},\
  }\href {https://doi.org/10.1103/PhysRevLett.110.260601} {\bibfield  {journal}
  {\bibinfo  {journal} {Phys. Rev. Lett.}\ }\textbf {\bibinfo {volume} {110}},\
  \bibinfo {pages} {260601} (\bibinfo {year} {2013}{\natexlab{a}})}\BibitemShut
  {NoStop}%
\bibitem [{\citenamefont {Serbyn}\ \emph
  {et~al.}(2013{\natexlab{b}})\citenamefont {Serbyn}, \citenamefont
  {Papi\'{c}},\ and\ \citenamefont {Abanin}}]{Serbyn2013Local}%
  \BibitemOpen
  \bibfield  {author} {\bibinfo {author} {\bibfnamefont {M.}~\bibnamefont
  {Serbyn}}, \bibinfo {author} {\bibfnamefont {Z.}~\bibnamefont {Papi\'{c}}},\
  and\ \bibinfo {author} {\bibfnamefont {D.~A.}\ \bibnamefont {Abanin}},\
  }\href {https://doi.org/10.1103/PhysRevLett.111.127201} {\bibfield  {journal}
  {\bibinfo  {journal} {Phys. Rev. Lett.}\ }\textbf {\bibinfo {volume} {111}},\
  \bibinfo {pages} {127201} (\bibinfo {year} {2013}{\natexlab{b}})}\BibitemShut
  {NoStop}%
\bibitem [{\citenamefont {Doggen}\ \emph {et~al.}(2018)\citenamefont {Doggen},
  \citenamefont {Schindler}, \citenamefont {Tikhonov}, \citenamefont {Mirlin},
  \citenamefont {Neupert}, \citenamefont {Polyakov},\ and\ \citenamefont
  {Gornyi}}]{Diggen2018Manybody}%
  \BibitemOpen
  \bibfield  {author} {\bibinfo {author} {\bibfnamefont {E.~V.~H.}\
  \bibnamefont {Doggen}}, \bibinfo {author} {\bibfnamefont {F.}~\bibnamefont
  {Schindler}}, \bibinfo {author} {\bibfnamefont {K.~S.}\ \bibnamefont
  {Tikhonov}}, \bibinfo {author} {\bibfnamefont {A.~D.}\ \bibnamefont
  {Mirlin}}, \bibinfo {author} {\bibfnamefont {T.}~\bibnamefont {Neupert}},
  \bibinfo {author} {\bibfnamefont {D.~G.}\ \bibnamefont {Polyakov}},\ and\
  \bibinfo {author} {\bibfnamefont {I.~V.}\ \bibnamefont {Gornyi}},\ }\href
  {https://doi.org/10.1103/PhysRevB.98.174202} {\bibfield  {journal} {\bibinfo
  {journal} {Phys. Rev. B}\ }\textbf {\bibinfo {volume} {98}},\ \bibinfo
  {pages} {174202} (\bibinfo {year} {2018})}\BibitemShut {NoStop}%
\bibitem [{\citenamefont {Luca}\ and\ \citenamefont
  {Scardicchio}(2013)}]{Deluca13}%
  \BibitemOpen
  \bibfield  {author} {\bibinfo {author} {\bibfnamefont {A.~D.}\ \bibnamefont
  {Luca}}\ and\ \bibinfo {author} {\bibfnamefont {A.}~\bibnamefont
  {Scardicchio}},\ }\href {https://doi.org/10.1209/0295-5075/101/37003}
  {\bibfield  {journal} {\bibinfo  {journal} {Europhys. Lett.}\ }\textbf
  {\bibinfo {volume} {101}},\ \bibinfo {pages} {37003} (\bibinfo {year}
  {2013})}\BibitemShut {NoStop}%
\bibitem [{\citenamefont {\v{S}untajs}\ \emph {et~al.}(2020)\citenamefont
  {\v{S}untajs}, \citenamefont {Bon\v{c}a}, \citenamefont {Prosen},\ and\
  \citenamefont {Vidmar}}]{Suntajs2020Quantum}%
  \BibitemOpen
  \bibfield  {author} {\bibinfo {author} {\bibfnamefont {J.}~\bibnamefont
  {\v{S}untajs}}, \bibinfo {author} {\bibfnamefont {J.}~\bibnamefont
  {Bon\v{c}a}}, \bibinfo {author} {\bibfnamefont {T.}~\bibnamefont {Prosen}},\
  and\ \bibinfo {author} {\bibfnamefont {L.}~\bibnamefont {Vidmar}},\ }\href
  {https://doi.org/10.1103/PhysRevE.102.062144} {\bibfield  {journal} {\bibinfo
   {journal} {Phys. Rev. E}\ }\textbf {\bibinfo {volume} {102}},\ \bibinfo
  {pages} {062144} (\bibinfo {year} {2020})}\BibitemShut {NoStop}%
\bibitem [{\citenamefont {Sels}\ and\ \citenamefont
  {Polkovnikov}(2021)}]{Sels2021Dynamical}%
  \BibitemOpen
  \bibfield  {author} {\bibinfo {author} {\bibfnamefont {D.}~\bibnamefont
  {Sels}}\ and\ \bibinfo {author} {\bibfnamefont {A.}~\bibnamefont
  {Polkovnikov}},\ }\href {https://doi.org/10.1103/PhysRevE.104.054105}
  {\bibfield  {journal} {\bibinfo  {journal} {Phys. Rev. E}\ }\textbf {\bibinfo
  {volume} {104}},\ \bibinfo {pages} {054105} (\bibinfo {year}
  {2021})}\BibitemShut {NoStop}%
\bibitem [{\citenamefont
  {Imbrie}(2016{\natexlab{a}})}]{Imbrie2016Diagonalization}%
  \BibitemOpen
  \bibfield  {author} {\bibinfo {author} {\bibfnamefont {J.~Z.}\ \bibnamefont
  {Imbrie}},\ }\href {https://doi.org/10.1103/PhysRevLett.117.027201}
  {\bibfield  {journal} {\bibinfo  {journal} {Phys. Rev. Lett.}\ }\textbf
  {\bibinfo {volume} {117}},\ \bibinfo {pages} {027201} (\bibinfo {year}
  {2016}{\natexlab{a}})}\BibitemShut {NoStop}%
\bibitem [{\citenamefont {Imbrie}(2016{\natexlab{b}})}]{Imbrie2016Many}%
  \BibitemOpen
  \bibfield  {author} {\bibinfo {author} {\bibfnamefont {J.~Z.}\ \bibnamefont
  {Imbrie}},\ }\href {https://doi.org/10.1007/s10955-016-1508-x} {\bibfield
  {journal} {\bibinfo  {journal} {J. Stat. Phys.}\ }\textbf {\bibinfo {volume}
  {163}},\ \bibinfo {pages} {998} (\bibinfo {year}
  {2016}{\natexlab{b}})}\BibitemShut {NoStop}%
\bibitem [{\citenamefont {De~Roeck}\ \emph {et~al.}(2024)\citenamefont
  {De~Roeck}, \citenamefont {Giacomin}, \citenamefont {Huveneers},\ and\
  \citenamefont {Prosniak}}]{deroeck2024absence}%
  \BibitemOpen
  \bibfield  {author} {\bibinfo {author} {\bibfnamefont {W.}~\bibnamefont
  {De~Roeck}}, \bibinfo {author} {\bibfnamefont {L.}~\bibnamefont {Giacomin}},
  \bibinfo {author} {\bibfnamefont {F.}~\bibnamefont {Huveneers}},\ and\
  \bibinfo {author} {\bibfnamefont {O.}~\bibnamefont {Prosniak}},\ }\href@noop
  {} {\bibinfo {title} {{Absence of Normal Heat Conduction in Strongly
  Disordered Interacting Quantum Chains}}} (\bibinfo {year} {2024}),\ \Eprint
  {https://arxiv.org/abs/2408.04338} {arXiv:2408.04338} \BibitemShut {NoStop}%
\bibitem [{\citenamefont {Chandran}\ \emph {et~al.}(2015)\citenamefont
  {Chandran}, \citenamefont {Laumann},\ and\ \citenamefont
  {Oganesyan}}]{Chandran2015Finite}%
  \BibitemOpen
  \bibfield  {author} {\bibinfo {author} {\bibfnamefont {A.}~\bibnamefont
  {Chandran}}, \bibinfo {author} {\bibfnamefont {C.~R.}\ \bibnamefont
  {Laumann}},\ and\ \bibinfo {author} {\bibfnamefont {V.}~\bibnamefont
  {Oganesyan}},\ }\href@noop {} {\bibinfo {title} {{Finite size scaling bounds
  on many-body localized phase transitions}}} (\bibinfo {year} {2015}),\
  \Eprint {https://arxiv.org/abs/1509.04285} {arXiv:1509.04285} \BibitemShut
  {NoStop}%
\bibitem [{\citenamefont {Panda}\ \emph {et~al.}(2020)\citenamefont {Panda},
  \citenamefont {Scardicchio}, \citenamefont {Schulz}, \citenamefont {Taylor},\
  and\ \citenamefont {Žnidarič}}]{panda2020can}%
  \BibitemOpen
  \bibfield  {author} {\bibinfo {author} {\bibfnamefont {R.~K.}\ \bibnamefont
  {Panda}}, \bibinfo {author} {\bibfnamefont {A.}~\bibnamefont {Scardicchio}},
  \bibinfo {author} {\bibfnamefont {M.}~\bibnamefont {Schulz}}, \bibinfo
  {author} {\bibfnamefont {S.~R.}\ \bibnamefont {Taylor}},\ and\ \bibinfo
  {author} {\bibfnamefont {M.}~\bibnamefont {Žnidarič}},\ }\href
  {https://doi.org/10.1209/0295-5075/128/67003} {\bibfield  {journal} {\bibinfo
   {journal} {Europhys. Lett.}\ }\textbf {\bibinfo {volume} {128}},\ \bibinfo
  {pages} {67003} (\bibinfo {year} {2020})}\BibitemShut {NoStop}%
\bibitem [{\citenamefont {Abanin}\ \emph {et~al.}(2021)\citenamefont {Abanin},
  \citenamefont {Bardarson}, \citenamefont {{De Tomasi}}, \citenamefont
  {Gopalakrishnan}, \citenamefont {Khemani}, \citenamefont {Parameswaran},
  \citenamefont {Pollmann}, \citenamefont {Potter}, \citenamefont {Serbyn},\
  and\ \citenamefont {Vasseur}}]{abanin2021distinguishing}%
  \BibitemOpen
  \bibfield  {author} {\bibinfo {author} {\bibfnamefont {D.}~\bibnamefont
  {Abanin}}, \bibinfo {author} {\bibfnamefont {J.}~\bibnamefont {Bardarson}},
  \bibinfo {author} {\bibfnamefont {G.}~\bibnamefont {{De Tomasi}}}, \bibinfo
  {author} {\bibfnamefont {S.}~\bibnamefont {Gopalakrishnan}}, \bibinfo
  {author} {\bibfnamefont {V.}~\bibnamefont {Khemani}}, \bibinfo {author}
  {\bibfnamefont {S.}~\bibnamefont {Parameswaran}}, \bibinfo {author}
  {\bibfnamefont {F.}~\bibnamefont {Pollmann}}, \bibinfo {author}
  {\bibfnamefont {A.}~\bibnamefont {Potter}}, \bibinfo {author} {\bibfnamefont
  {M.}~\bibnamefont {Serbyn}},\ and\ \bibinfo {author} {\bibfnamefont
  {R.}~\bibnamefont {Vasseur}},\ }\href
  {https://doi.org/10.1016/j.aop.2021.168415} {\bibfield  {journal} {\bibinfo
  {journal} {Ann. Phys. (Amsterdam)}\ }\textbf {\bibinfo {volume} {427}},\
  \bibinfo {pages} {168415} (\bibinfo {year} {2021})}\BibitemShut {NoStop}%
\bibitem [{\citenamefont {Sierant}\ and\ \citenamefont
  {Zakrzewski}(2022)}]{sierant2022challenges}%
  \BibitemOpen
  \bibfield  {author} {\bibinfo {author} {\bibfnamefont {P.}~\bibnamefont
  {Sierant}}\ and\ \bibinfo {author} {\bibfnamefont {J.}~\bibnamefont
  {Zakrzewski}},\ }\href {https://doi.org/10.1103/PhysRevB.105.224203}
  {\bibfield  {journal} {\bibinfo  {journal} {Phys. Rev. B}\ }\textbf {\bibinfo
  {volume} {105}},\ \bibinfo {pages} {224203} (\bibinfo {year}
  {2022})}\BibitemShut {NoStop}%
\bibitem [{\citenamefont {Sierant}\ \emph
  {et~al.}(2023{\natexlab{a}})\citenamefont {Sierant}, \citenamefont
  {Lewenstein}, \citenamefont {Scardicchio},\ and\ \citenamefont
  {Zakrzewski}}]{sierant2023stability}%
  \BibitemOpen
  \bibfield  {author} {\bibinfo {author} {\bibfnamefont {P.}~\bibnamefont
  {Sierant}}, \bibinfo {author} {\bibfnamefont {M.}~\bibnamefont {Lewenstein}},
  \bibinfo {author} {\bibfnamefont {A.}~\bibnamefont {Scardicchio}},\ and\
  \bibinfo {author} {\bibfnamefont {J.}~\bibnamefont {Zakrzewski}},\ }\href
  {https://doi.org/10.1103/PhysRevB.107.115132} {\bibfield  {journal} {\bibinfo
   {journal} {Phys. Rev. B}\ }\textbf {\bibinfo {volume} {107}},\ \bibinfo
  {pages} {115132} (\bibinfo {year} {2023}{\natexlab{a}})}\BibitemShut
  {NoStop}%
\bibitem [{\citenamefont {Sierant}\ \emph {et~al.}(2025)\citenamefont
  {Sierant}, \citenamefont {Lewenstein}, \citenamefont {Scardicchio},
  \citenamefont {Vidmar},\ and\ \citenamefont
  {Zakrzewski}}]{sierant24MBLreview}%
  \BibitemOpen
  \bibfield  {author} {\bibinfo {author} {\bibfnamefont {P.}~\bibnamefont
  {Sierant}}, \bibinfo {author} {\bibfnamefont {M.}~\bibnamefont {Lewenstein}},
  \bibinfo {author} {\bibfnamefont {A.}~\bibnamefont {Scardicchio}}, \bibinfo
  {author} {\bibfnamefont {L.}~\bibnamefont {Vidmar}},\ and\ \bibinfo {author}
  {\bibfnamefont {J.}~\bibnamefont {Zakrzewski}},\ }\href
  {https://doi.org/10.1088/1361-6633/ad9756} {\bibfield  {journal} {\bibinfo
  {journal} {Rep. Progr. Phys.}\ }\textbf {\bibinfo {volume} {88}},\ \bibinfo
  {pages} {026502} (\bibinfo {year} {2025})}\BibitemShut {NoStop}%
\bibitem [{\citenamefont {Luitz}\ and\ \citenamefont
  {Lev}(2017)}]{Luitz2017Ergodic}%
  \BibitemOpen
  \bibfield  {author} {\bibinfo {author} {\bibfnamefont {D.~J.}\ \bibnamefont
  {Luitz}}\ and\ \bibinfo {author} {\bibfnamefont {Y.~B.}\ \bibnamefont
  {Lev}},\ }\href {https://doi.org/10.1002/andp.201600350} {\bibfield
  {journal} {\bibinfo  {journal} {Ann. Phys. (Berlin)}\ }\textbf {\bibinfo
  {volume} {529}},\ \bibinfo {pages} {1600350} (\bibinfo {year}
  {2017})}\BibitemShut {NoStop}%
\bibitem [{\citenamefont {Crowley}\ and\ \citenamefont
  {Chandran}(2022)}]{Crowley2022Constructive}%
  \BibitemOpen
  \bibfield  {author} {\bibinfo {author} {\bibfnamefont {P.~J.~D.}\
  \bibnamefont {Crowley}}\ and\ \bibinfo {author} {\bibfnamefont
  {A.}~\bibnamefont {Chandran}},\ }\href
  {https://doi.org/10.21468/SciPostPhys.12.6.201} {\bibfield  {journal}
  {\bibinfo  {journal} {SciPost Phys.}\ }\textbf {\bibinfo {volume} {12}},\
  \bibinfo {pages} {201} (\bibinfo {year} {2022})}\BibitemShut {NoStop}%
\bibitem [{\citenamefont {Morningstar}\ \emph {et~al.}(2022)\citenamefont
  {Morningstar}, \citenamefont {Colmenarez}, \citenamefont {Khemani},
  \citenamefont {Luitz},\ and\ \citenamefont
  {Huse}}]{Morningstar2022Avalanches}%
  \BibitemOpen
  \bibfield  {author} {\bibinfo {author} {\bibfnamefont {A.}~\bibnamefont
  {Morningstar}}, \bibinfo {author} {\bibfnamefont {L.}~\bibnamefont
  {Colmenarez}}, \bibinfo {author} {\bibfnamefont {V.}~\bibnamefont {Khemani}},
  \bibinfo {author} {\bibfnamefont {D.~J.}\ \bibnamefont {Luitz}},\ and\
  \bibinfo {author} {\bibfnamefont {D.~A.}\ \bibnamefont {Huse}},\ }\href
  {https://doi.org/10.1103/PhysRevB.105.174205} {\bibfield  {journal} {\bibinfo
   {journal} {Phys. Rev. B}\ }\textbf {\bibinfo {volume} {105}},\ \bibinfo
  {pages} {174205} (\bibinfo {year} {2022})}\BibitemShut {NoStop}%
\bibitem [{\citenamefont {Abrahams}\ \emph {et~al.}(1979)\citenamefont
  {Abrahams}, \citenamefont {Anderson}, \citenamefont {Licciardello},\ and\
  \citenamefont {Ramakrishnan}}]{abrahams1979scaling}%
  \BibitemOpen
  \bibfield  {author} {\bibinfo {author} {\bibfnamefont {E.}~\bibnamefont
  {Abrahams}}, \bibinfo {author} {\bibfnamefont {P.~W.}\ \bibnamefont
  {Anderson}}, \bibinfo {author} {\bibfnamefont {D.~C.}\ \bibnamefont
  {Licciardello}},\ and\ \bibinfo {author} {\bibfnamefont {T.~V.}\ \bibnamefont
  {Ramakrishnan}},\ }\href {https://doi.org/10.1103/PhysRevLett.42.673}
  {\bibfield  {journal} {\bibinfo  {journal} {Phys. Rev. Lett.}\ }\textbf
  {\bibinfo {volume} {42}},\ \bibinfo {pages} {673} (\bibinfo {year}
  {1979})}\BibitemShut {NoStop}%
\bibitem [{\citenamefont {Vosk}\ and\ \citenamefont
  {Altman}(2013)}]{Vosk2013Many}%
  \BibitemOpen
  \bibfield  {author} {\bibinfo {author} {\bibfnamefont {R.}~\bibnamefont
  {Vosk}}\ and\ \bibinfo {author} {\bibfnamefont {E.}~\bibnamefont {Altman}},\
  }\href {https://doi.org/10.1103/PhysRevLett.110.067204} {\bibfield  {journal}
  {\bibinfo  {journal} {Phys. Rev. Lett.}\ }\textbf {\bibinfo {volume} {110}},\
  \bibinfo {pages} {067204} (\bibinfo {year} {2013})}\BibitemShut {NoStop}%
\bibitem [{\citenamefont {Potter}\ \emph {et~al.}(2015)\citenamefont {Potter},
  \citenamefont {Vasseur},\ and\ \citenamefont
  {Parameswaran}}]{Potter2015Universal}%
  \BibitemOpen
  \bibfield  {author} {\bibinfo {author} {\bibfnamefont {A.~C.}\ \bibnamefont
  {Potter}}, \bibinfo {author} {\bibfnamefont {R.}~\bibnamefont {Vasseur}},\
  and\ \bibinfo {author} {\bibfnamefont {S.~A.}\ \bibnamefont {Parameswaran}},\
  }\href {https://doi.org/10.1103/PhysRevX.5.031033} {\bibfield  {journal}
  {\bibinfo  {journal} {Phys. Rev. X}\ }\textbf {\bibinfo {volume} {5}},\
  \bibinfo {pages} {031033} (\bibinfo {year} {2015})}\BibitemShut {NoStop}%
\bibitem [{\citenamefont {Zhang}\ \emph {et~al.}(2016)\citenamefont {Zhang},
  \citenamefont {Zhao}, \citenamefont {Devakul},\ and\ \citenamefont
  {Huse}}]{Zhang2016Many}%
  \BibitemOpen
  \bibfield  {author} {\bibinfo {author} {\bibfnamefont {L.}~\bibnamefont
  {Zhang}}, \bibinfo {author} {\bibfnamefont {B.}~\bibnamefont {Zhao}},
  \bibinfo {author} {\bibfnamefont {T.}~\bibnamefont {Devakul}},\ and\ \bibinfo
  {author} {\bibfnamefont {D.~A.}\ \bibnamefont {Huse}},\ }\href
  {https://doi.org/10.1103/PhysRevB.93.224201} {\bibfield  {journal} {\bibinfo
  {journal} {Phys. Rev. B}\ }\textbf {\bibinfo {volume} {93}},\ \bibinfo
  {pages} {224201} (\bibinfo {year} {2016})}\BibitemShut {NoStop}%
\bibitem [{\citenamefont {Thiery}\ \emph {et~al.}(2017)\citenamefont {Thiery},
  \citenamefont {M\"uller},\ and\ \citenamefont
  {De~Roeck}}]{Thiery2017Microscopically}%
  \BibitemOpen
  \bibfield  {author} {\bibinfo {author} {\bibfnamefont {T.}~\bibnamefont
  {Thiery}}, \bibinfo {author} {\bibfnamefont {M.}~\bibnamefont {M\"uller}},\
  and\ \bibinfo {author} {\bibfnamefont {W.}~\bibnamefont {De~Roeck}},\
  }\href@noop {} {\bibinfo {title} {{A microscopically motivated
  renormalization scheme for the MBL/ETH transition}}} (\bibinfo {year}
  {2017}),\ \Eprint {https://arxiv.org/abs/1711.09880} {arXiv:1711.09880}
  \BibitemShut {NoStop}%
\bibitem [{\citenamefont {Goremykina}\ \emph {et~al.}(2019)\citenamefont
  {Goremykina}, \citenamefont {Vasseur},\ and\ \citenamefont
  {Serbyn}}]{Goremykina2019Analytically}%
  \BibitemOpen
  \bibfield  {author} {\bibinfo {author} {\bibfnamefont {A.}~\bibnamefont
  {Goremykina}}, \bibinfo {author} {\bibfnamefont {R.}~\bibnamefont
  {Vasseur}},\ and\ \bibinfo {author} {\bibfnamefont {M.}~\bibnamefont
  {Serbyn}},\ }\href {https://doi.org/10.1103/PhysRevLett.122.040601}
  {\bibfield  {journal} {\bibinfo  {journal} {Phys. Rev. Lett.}\ }\textbf
  {\bibinfo {volume} {122}},\ \bibinfo {pages} {040601} (\bibinfo {year}
  {2019})}\BibitemShut {NoStop}%
\bibitem [{\citenamefont {Dumitrescu}\ \emph {et~al.}(2019)\citenamefont
  {Dumitrescu}, \citenamefont {Goremykina}, \citenamefont {Parameswaran},
  \citenamefont {Serbyn},\ and\ \citenamefont
  {Vasseur}}]{Dumitrescu2019Kosterlitz}%
  \BibitemOpen
  \bibfield  {author} {\bibinfo {author} {\bibfnamefont {P.~T.}\ \bibnamefont
  {Dumitrescu}}, \bibinfo {author} {\bibfnamefont {A.}~\bibnamefont
  {Goremykina}}, \bibinfo {author} {\bibfnamefont {S.~A.}\ \bibnamefont
  {Parameswaran}}, \bibinfo {author} {\bibfnamefont {M.}~\bibnamefont
  {Serbyn}},\ and\ \bibinfo {author} {\bibfnamefont {R.}~\bibnamefont
  {Vasseur}},\ }\href {https://doi.org/10.1103/PhysRevB.99.094205} {\bibfield
  {journal} {\bibinfo  {journal} {Phys. Rev. B}\ }\textbf {\bibinfo {volume}
  {99}},\ \bibinfo {pages} {094205} (\bibinfo {year} {2019})}\BibitemShut
  {NoStop}%
\bibitem [{\citenamefont {Morningstar}\ and\ \citenamefont
  {Huse}(2019)}]{Morningstar2019Renormalization}%
  \BibitemOpen
  \bibfield  {author} {\bibinfo {author} {\bibfnamefont {A.}~\bibnamefont
  {Morningstar}}\ and\ \bibinfo {author} {\bibfnamefont {D.~A.}\ \bibnamefont
  {Huse}},\ }\href {https://doi.org/10.1103/PhysRevB.99.224205} {\bibfield
  {journal} {\bibinfo  {journal} {Phys. Rev. B}\ }\textbf {\bibinfo {volume}
  {99}},\ \bibinfo {pages} {224205} (\bibinfo {year} {2019})}\BibitemShut
  {NoStop}%
\bibitem [{\citenamefont {Morningstar}\ \emph {et~al.}(2020)\citenamefont
  {Morningstar}, \citenamefont {Huse},\ and\ \citenamefont
  {Imbrie}}]{Morningstar2020Many}%
  \BibitemOpen
  \bibfield  {author} {\bibinfo {author} {\bibfnamefont {A.}~\bibnamefont
  {Morningstar}}, \bibinfo {author} {\bibfnamefont {D.~A.}\ \bibnamefont
  {Huse}},\ and\ \bibinfo {author} {\bibfnamefont {J.~Z.}\ \bibnamefont
  {Imbrie}},\ }\href {https://doi.org/10.1103/PhysRevB.102.125134} {\bibfield
  {journal} {\bibinfo  {journal} {Phys. Rev. B}\ }\textbf {\bibinfo {volume}
  {102}},\ \bibinfo {pages} {125134} (\bibinfo {year} {2020})}\BibitemShut
  {NoStop}%
\bibitem [{\citenamefont {Ros}\ \emph {et~al.}(2015)\citenamefont {Ros},
  \citenamefont {Müller},\ and\ \citenamefont
  {Scardicchio}}]{ros2015integrals}%
  \BibitemOpen
  \bibfield  {author} {\bibinfo {author} {\bibfnamefont {V.}~\bibnamefont
  {Ros}}, \bibinfo {author} {\bibfnamefont {M.}~\bibnamefont {Müller}},\ and\
  \bibinfo {author} {\bibfnamefont {A.}~\bibnamefont {Scardicchio}},\ }\href
  {https://doi.org/10.1016/j.nuclphysb.2014.12.014} {\bibfield  {journal}
  {\bibinfo  {journal} {Nucl. Phys. B}\ }\textbf {\bibinfo {volume} {891}},\
  \bibinfo {pages} {420} (\bibinfo {year} {2015})}\BibitemShut {NoStop}%
\bibitem [{\citenamefont {De~Luca}\ \emph {et~al.}(2014)\citenamefont
  {De~Luca}, \citenamefont {Altshuler}, \citenamefont {Kravtsov},\ and\
  \citenamefont {Scardicchio}}]{de2014anderson}%
  \BibitemOpen
  \bibfield  {author} {\bibinfo {author} {\bibfnamefont {A.}~\bibnamefont
  {De~Luca}}, \bibinfo {author} {\bibfnamefont {B.~L.}\ \bibnamefont
  {Altshuler}}, \bibinfo {author} {\bibfnamefont {V.~E.}\ \bibnamefont
  {Kravtsov}},\ and\ \bibinfo {author} {\bibfnamefont {A.}~\bibnamefont
  {Scardicchio}},\ }\href {https://doi.org/10.1103/PhysRevLett.113.046806}
  {\bibfield  {journal} {\bibinfo  {journal} {Phys. Rev. Lett.}\ }\textbf
  {\bibinfo {volume} {113}},\ \bibinfo {pages} {046806} (\bibinfo {year}
  {2014})}\BibitemShut {NoStop}%
\bibitem [{\citenamefont {Tikhonov}\ \emph {et~al.}(2016)\citenamefont
  {Tikhonov}, \citenamefont {Mirlin},\ and\ \citenamefont
  {Skvortsov}}]{tikhonov2016Anderson}%
  \BibitemOpen
  \bibfield  {author} {\bibinfo {author} {\bibfnamefont {K.~S.}\ \bibnamefont
  {Tikhonov}}, \bibinfo {author} {\bibfnamefont {A.~D.}\ \bibnamefont
  {Mirlin}},\ and\ \bibinfo {author} {\bibfnamefont {M.~A.}\ \bibnamefont
  {Skvortsov}},\ }\href {https://doi.org/10.1103/PhysRevB.94.220203} {\bibfield
   {journal} {\bibinfo  {journal} {Phys. Rev. B}\ }\textbf {\bibinfo {volume}
  {94}},\ \bibinfo {pages} {220203} (\bibinfo {year} {2016})}\BibitemShut
  {NoStop}%
\bibitem [{\citenamefont {Bera}\ \emph {et~al.}(2018)\citenamefont {Bera},
  \citenamefont {De~Tomasi}, \citenamefont {Khaymovich},\ and\ \citenamefont
  {Scardicchio}}]{bera2018return}%
  \BibitemOpen
  \bibfield  {author} {\bibinfo {author} {\bibfnamefont {S.}~\bibnamefont
  {Bera}}, \bibinfo {author} {\bibfnamefont {G.}~\bibnamefont {De~Tomasi}},
  \bibinfo {author} {\bibfnamefont {I.~M.}\ \bibnamefont {Khaymovich}},\ and\
  \bibinfo {author} {\bibfnamefont {A.}~\bibnamefont {Scardicchio}},\ }\href
  {https://doi.org/10.1103/PhysRevB.98.134205} {\bibfield  {journal} {\bibinfo
  {journal} {Phys. Rev. B}\ }\textbf {\bibinfo {volume} {98}},\ \bibinfo
  {pages} {134205} (\bibinfo {year} {2018})}\BibitemShut {NoStop}%
\bibitem [{\citenamefont {Kravtsov}\ \emph {et~al.}(2018)\citenamefont
  {Kravtsov}, \citenamefont {Altshuler},\ and\ \citenamefont
  {Ioffe}}]{kravtsov2018non}%
  \BibitemOpen
  \bibfield  {author} {\bibinfo {author} {\bibfnamefont {V.}~\bibnamefont
  {Kravtsov}}, \bibinfo {author} {\bibfnamefont {B.}~\bibnamefont
  {Altshuler}},\ and\ \bibinfo {author} {\bibfnamefont {L.}~\bibnamefont
  {Ioffe}},\ }\href {https://doi.org/10.1016/j.aop.2017.12.009} {\bibfield
  {journal} {\bibinfo  {journal} {Ann. Phys. (Amsterdam)}\ }\textbf {\bibinfo
  {volume} {389}},\ \bibinfo {pages} {148} (\bibinfo {year}
  {2018})}\BibitemShut {NoStop}%
\bibitem [{\citenamefont {Parisi}\ \emph {et~al.}(2019)\citenamefont {Parisi},
  \citenamefont {Pascazio}, \citenamefont {Pietracaprina}, \citenamefont
  {Ros},\ and\ \citenamefont {Scardicchio}}]{parisi2019anderson}%
  \BibitemOpen
  \bibfield  {author} {\bibinfo {author} {\bibfnamefont {G.}~\bibnamefont
  {Parisi}}, \bibinfo {author} {\bibfnamefont {S.}~\bibnamefont {Pascazio}},
  \bibinfo {author} {\bibfnamefont {F.}~\bibnamefont {Pietracaprina}}, \bibinfo
  {author} {\bibfnamefont {V.}~\bibnamefont {Ros}},\ and\ \bibinfo {author}
  {\bibfnamefont {A.}~\bibnamefont {Scardicchio}},\ }\href
  {https://doi.org/10.1088/1751-8121/ab56e8} {\bibfield  {journal} {\bibinfo
  {journal} {J. Phys. A}\ }\textbf {\bibinfo {volume} {53}},\ \bibinfo {pages}
  {014003} (\bibinfo {year} {2019})}\BibitemShut {NoStop}%
\bibitem [{\citenamefont {Tikhonov}\ and\ \citenamefont
  {Mirlin}(2021)}]{tikhonov2021AndersonMBL}%
  \BibitemOpen
  \bibfield  {author} {\bibinfo {author} {\bibfnamefont {K.}~\bibnamefont
  {Tikhonov}}\ and\ \bibinfo {author} {\bibfnamefont {A.}~\bibnamefont
  {Mirlin}},\ }\href {https://doi.org/10.1016/j.aop.2021.168525} {\bibfield
  {journal} {\bibinfo  {journal} {Ann. Phys. (Amsterdam)}\ }\textbf {\bibinfo
  {volume} {435}},\ \bibinfo {pages} {168525} (\bibinfo {year}
  {2021})}\BibitemShut {NoStop}%
\bibitem [{\citenamefont {Sierant}\ \emph
  {et~al.}(2023{\natexlab{b}})\citenamefont {Sierant}, \citenamefont
  {Lewenstein},\ and\ \citenamefont {Scardicchio}}]{sierant2023universality}%
  \BibitemOpen
  \bibfield  {author} {\bibinfo {author} {\bibfnamefont {P.}~\bibnamefont
  {Sierant}}, \bibinfo {author} {\bibfnamefont {M.}~\bibnamefont
  {Lewenstein}},\ and\ \bibinfo {author} {\bibfnamefont {A.}~\bibnamefont
  {Scardicchio}},\ }\href {https://doi.org/10.21468/SciPostPhys.15.2.045}
  {\bibfield  {journal} {\bibinfo  {journal} {SciPost Phys.}\ }\textbf
  {\bibinfo {volume} {15}},\ \bibinfo {pages} {045} (\bibinfo {year}
  {2023}{\natexlab{b}})}\BibitemShut {NoStop}%
\bibitem [{\citenamefont {Vanoni}\ and\ \citenamefont
  {Vitale}(2024)}]{vanoni2023analysis}%
  \BibitemOpen
  \bibfield  {author} {\bibinfo {author} {\bibfnamefont {C.}~\bibnamefont
  {Vanoni}}\ and\ \bibinfo {author} {\bibfnamefont {V.}~\bibnamefont
  {Vitale}},\ }\href {https://doi.org/10.1103/PhysRevB.110.024204} {\bibfield
  {journal} {\bibinfo  {journal} {Phys. Rev. B}\ }\textbf {\bibinfo {volume}
  {110}},\ \bibinfo {pages} {024204} (\bibinfo {year} {2024})}\BibitemShut
  {NoStop}%
\bibitem [{\citenamefont {Vanoni}\ \emph {et~al.}(2024)\citenamefont {Vanoni},
  \citenamefont {Altshuler}, \citenamefont {Kravtsov},\ and\ \citenamefont
  {Scardicchio}}]{vanoni2023renormalization}%
  \BibitemOpen
  \bibfield  {author} {\bibinfo {author} {\bibfnamefont {C.}~\bibnamefont
  {Vanoni}}, \bibinfo {author} {\bibfnamefont {B.~L.}\ \bibnamefont
  {Altshuler}}, \bibinfo {author} {\bibfnamefont {V.~E.}\ \bibnamefont
  {Kravtsov}},\ and\ \bibinfo {author} {\bibfnamefont {A.}~\bibnamefont
  {Scardicchio}},\ }\bibfield  {journal} {\bibinfo  {journal} {Proc. Natl.
  Acad. Sci. (USA)}\ }\textbf {\bibinfo {volume} {121}},\ \href
  {https://doi.org/10.1073/pnas.2401955121} {10.1073/pnas.2401955121} (\bibinfo
  {year} {2024})\BibitemShut {NoStop}%
\bibitem [{\citenamefont {Berezinskii}(1971)}]{Berezinskii1971Destruction}%
  \BibitemOpen
  \bibfield  {author} {\bibinfo {author} {\bibfnamefont {V.~L.}\ \bibnamefont
  {Berezinskii}},\ }\href@noop {} {\bibfield  {journal} {\bibinfo  {journal}
  {Sov. Phys. JETP}\ }\textbf {\bibinfo {volume} {32}},\ \bibinfo {pages} {493}
  (\bibinfo {year} {1971})}\BibitemShut {NoStop}%
\bibitem [{\citenamefont {Berezinskii}(1972)}]{Berezinskii1972Destruction}%
  \BibitemOpen
  \bibfield  {author} {\bibinfo {author} {\bibfnamefont {V.~L.}\ \bibnamefont
  {Berezinskii}},\ }\href@noop {} {\bibfield  {journal} {\bibinfo  {journal}
  {Sov. Phys. JETP}\ }\textbf {\bibinfo {volume} {34}},\ \bibinfo {pages} {610}
  (\bibinfo {year} {1972})}\BibitemShut {NoStop}%
\bibitem [{\citenamefont {Kosterlitz}\ and\ \citenamefont
  {Thouless}(1973)}]{Kosterlitz1973Ordering}%
  \BibitemOpen
  \bibfield  {author} {\bibinfo {author} {\bibfnamefont {J.~M.}\ \bibnamefont
  {Kosterlitz}}\ and\ \bibinfo {author} {\bibfnamefont {D.~J.}\ \bibnamefont
  {Thouless}},\ }\href {https://doi.org/10.1088/0022-3719/6/7/010} {\bibfield
  {journal} {\bibinfo  {journal} {J. Phys. C}\ }\textbf {\bibinfo {volume}
  {6}},\ \bibinfo {pages} {1181} (\bibinfo {year} {1973})}\BibitemShut
  {NoStop}%
\bibitem [{\citenamefont {Altshuler}\ \emph {et~al.}(2024)\citenamefont
  {Altshuler}, \citenamefont {Kravtsov}, \citenamefont {Scardicchio},
  \citenamefont {Sierant},\ and\ \citenamefont
  {Vanoni}}]{altshuler2024renormalization}%
  \BibitemOpen
  \bibfield  {author} {\bibinfo {author} {\bibfnamefont {B.~L.}\ \bibnamefont
  {Altshuler}}, \bibinfo {author} {\bibfnamefont {V.~E.}\ \bibnamefont
  {Kravtsov}}, \bibinfo {author} {\bibfnamefont {A.}~\bibnamefont
  {Scardicchio}}, \bibinfo {author} {\bibfnamefont {P.}~\bibnamefont
  {Sierant}},\ and\ \bibinfo {author} {\bibfnamefont {C.}~\bibnamefont
  {Vanoni}},\ }\href {https://arxiv.org/abs/2403.01974} {\bibfield  {journal}
  {\bibinfo  {journal} {arXiv:2403.01974}\ } (\bibinfo {year}
  {2024})}\BibitemShut {NoStop}%
\bibitem [{\citenamefont {Laumann}\ \emph {et~al.}(2014)\citenamefont
  {Laumann}, \citenamefont {Pal},\ and\ \citenamefont
  {Scardicchio}}]{Laumann2014MBMobility}%
  \BibitemOpen
  \bibfield  {author} {\bibinfo {author} {\bibfnamefont {C.~R.}\ \bibnamefont
  {Laumann}}, \bibinfo {author} {\bibfnamefont {A.}~\bibnamefont {Pal}},\ and\
  \bibinfo {author} {\bibfnamefont {A.}~\bibnamefont {Scardicchio}},\ }\href
  {https://doi.org/10.1103/PhysRevLett.113.200405} {\bibfield  {journal}
  {\bibinfo  {journal} {Phys. Rev. Lett.}\ }\textbf {\bibinfo {volume} {113}},\
  \bibinfo {pages} {200405} (\bibinfo {year} {2014})}\BibitemShut {NoStop}%
\bibitem [{\citenamefont {Baldwin}\ \emph {et~al.}(2016)\citenamefont
  {Baldwin}, \citenamefont {Laumann}, \citenamefont {Pal},\ and\ \citenamefont
  {Scardicchio}}]{Baldwin2016ManyBody}%
  \BibitemOpen
  \bibfield  {author} {\bibinfo {author} {\bibfnamefont {C.~L.}\ \bibnamefont
  {Baldwin}}, \bibinfo {author} {\bibfnamefont {C.~R.}\ \bibnamefont
  {Laumann}}, \bibinfo {author} {\bibfnamefont {A.}~\bibnamefont {Pal}},\ and\
  \bibinfo {author} {\bibfnamefont {A.}~\bibnamefont {Scardicchio}},\ }\href
  {https://doi.org/10.1103/PhysRevB.93.024202} {\bibfield  {journal} {\bibinfo
  {journal} {Phys. Rev. B}\ }\textbf {\bibinfo {volume} {93}},\ \bibinfo
  {pages} {024202} (\bibinfo {year} {2016})}\BibitemShut {NoStop}%
\bibitem [{\citenamefont {Kutlin}\ and\ \citenamefont
  {Vanoni}(2025)}]{kutlin2024investigating}%
  \BibitemOpen
  \bibfield  {author} {\bibinfo {author} {\bibfnamefont {A.}~\bibnamefont
  {Kutlin}}\ and\ \bibinfo {author} {\bibfnamefont {C.}~\bibnamefont
  {Vanoni}},\ }\href {https://doi.org/10.21468/SciPostPhys.18.3.090} {\bibfield
   {journal} {\bibinfo  {journal} {SciPost Phys.}\ }\textbf {\bibinfo {volume}
  {18}},\ \bibinfo {pages} {090} (\bibinfo {year} {2025})}\BibitemShut
  {NoStop}%
\bibitem [{\citenamefont {Niedda}\ \emph {et~al.}(2024)\citenamefont {Niedda},
  \citenamefont {Bracci-Testasecca}, \citenamefont {Magnifico}, \citenamefont
  {Balducci}, \citenamefont {Vanoni},\ and\ \citenamefont
  {Scardicchio}}]{Niedda2024}%
  \BibitemOpen
  \bibfield  {author} {\bibinfo {author} {\bibfnamefont {J.}~\bibnamefont
  {Niedda}}, \bibinfo {author} {\bibfnamefont {G.}~\bibnamefont
  {Bracci-Testasecca}}, \bibinfo {author} {\bibfnamefont {G.}~\bibnamefont
  {Magnifico}}, \bibinfo {author} {\bibfnamefont {F.}~\bibnamefont {Balducci}},
  \bibinfo {author} {\bibfnamefont {C.}~\bibnamefont {Vanoni}},\ and\ \bibinfo
  {author} {\bibfnamefont {A.}~\bibnamefont {Scardicchio}},\ }\href
  {https://arxiv.org/abs/2410.12430} {\bibfield  {journal} {\bibinfo  {journal}
  {arXiv:2410.12430}\ } (\bibinfo {year} {2024})}\BibitemShut {NoStop}%
\bibitem [{\citenamefont {Świętek}\ \emph {et~al.}(2024)\citenamefont
  {Świętek}, \citenamefont {Hopjan}, \citenamefont {Vanoni}, \citenamefont
  {Scardicchio},\ and\ \citenamefont {Vidmar}}]{Swietek_2024}%
  \BibitemOpen
  \bibfield  {author} {\bibinfo {author} {\bibfnamefont {R.}~\bibnamefont
  {Świętek}}, \bibinfo {author} {\bibfnamefont {M.}~\bibnamefont {Hopjan}},
  \bibinfo {author} {\bibfnamefont {C.}~\bibnamefont {Vanoni}}, \bibinfo
  {author} {\bibfnamefont {A.}~\bibnamefont {Scardicchio}},\ and\ \bibinfo
  {author} {\bibfnamefont {L.}~\bibnamefont {Vidmar}},\ }\href
  {https://arxiv.org/abs/2412.15331} {\bibfield  {journal} {\bibinfo  {journal}
  {arXiv:2412.15331}\ } (\bibinfo {year} {2024})}\BibitemShut {NoStop}%
\bibitem [{\citenamefont {Derrida}(1980)}]{Derrida1980REM}%
  \BibitemOpen
  \bibfield  {author} {\bibinfo {author} {\bibfnamefont {B.}~\bibnamefont
  {Derrida}},\ }\href {https://doi.org/10.1103/PhysRevLett.45.79} {\bibfield
  {journal} {\bibinfo  {journal} {Phys. Rev. Lett.}\ }\textbf {\bibinfo
  {volume} {45}},\ \bibinfo {pages} {79} (\bibinfo {year} {1980})}\BibitemShut
  {NoStop}%
\bibitem [{\citenamefont {Derrida}(1981)}]{Derrida1981REM}%
  \BibitemOpen
  \bibfield  {author} {\bibinfo {author} {\bibfnamefont {B.}~\bibnamefont
  {Derrida}},\ }\href {https://doi.org/10.1103/PhysRevB.24.2613} {\bibfield
  {journal} {\bibinfo  {journal} {Phys. Rev. B}\ }\textbf {\bibinfo {volume}
  {24}},\ \bibinfo {pages} {2613} (\bibinfo {year} {1981})}\BibitemShut
  {NoStop}%
\bibitem [{\citenamefont {Gross}\ and\ \citenamefont
  {Mezard}(1984)}]{GrossMezard1984}%
  \BibitemOpen
  \bibfield  {author} {\bibinfo {author} {\bibfnamefont {D.}~\bibnamefont
  {Gross}}\ and\ \bibinfo {author} {\bibfnamefont {M.}~\bibnamefont {Mezard}},\
  }\href {https://doi.org/10.1016/0550-3213(84)90237-2} {\bibfield  {journal}
  {\bibinfo  {journal} {Nucl. Phys. B}\ }\textbf {\bibinfo {volume} {240}},\
  \bibinfo {pages} {431} (\bibinfo {year} {1984})}\BibitemShut {NoStop}%
\bibitem [{\citenamefont {Goldschmidt}(1990)}]{Goldschmidt_1990}%
  \BibitemOpen
  \bibfield  {author} {\bibinfo {author} {\bibfnamefont {Y.~Y.}\ \bibnamefont
  {Goldschmidt}},\ }\href {https://doi.org/10.1103/PhysRevB.41.4858} {\bibfield
   {journal} {\bibinfo  {journal} {Phys. Rev. B}\ }\textbf {\bibinfo {volume}
  {41}},\ \bibinfo {pages} {4858} (\bibinfo {year} {1990})}\BibitemShut
  {NoStop}%
\bibitem [{\citenamefont {Manai}\ and\ \citenamefont
  {Warzel}(2020)}]{Manai2020Phase}%
  \BibitemOpen
  \bibfield  {author} {\bibinfo {author} {\bibfnamefont {C.}~\bibnamefont
  {Manai}}\ and\ \bibinfo {author} {\bibfnamefont {S.}~\bibnamefont {Warzel}},\
  }\href {https://doi.org/10.1007/s10955-020-02492-5} {\bibfield  {journal}
  {\bibinfo  {journal} {J. Stat. Phys.}\ }\textbf {\bibinfo {volume} {180}},\
  \bibinfo {pages} {654} (\bibinfo {year} {2020})}\BibitemShut {NoStop}%
\bibitem [{\citenamefont {Manai}\ and\ \citenamefont
  {Warzel}(2023)}]{Manai2023Spectral}%
  \BibitemOpen
  \bibfield  {author} {\bibinfo {author} {\bibfnamefont {C.}~\bibnamefont
  {Manai}}\ and\ \bibinfo {author} {\bibfnamefont {S.}~\bibnamefont {Warzel}},\
  }\href {https://doi.org/10.1007/s00220-023-04743-4} {\bibfield  {journal}
  {\bibinfo  {journal} {Commun. Math. Phys.}\ }\textbf {\bibinfo {volume}
  {402}},\ \bibinfo {pages} {1259} (\bibinfo {year} {2023})}\BibitemShut
  {NoStop}%
\bibitem [{\citenamefont {Biroli}\ \emph {et~al.}(2021)\citenamefont {Biroli},
  \citenamefont {Facoetti}, \citenamefont {Schir\'o}, \citenamefont {Tarzia},\
  and\ \citenamefont {Vivo}}]{Biroli2021Out}%
  \BibitemOpen
  \bibfield  {author} {\bibinfo {author} {\bibfnamefont {G.}~\bibnamefont
  {Biroli}}, \bibinfo {author} {\bibfnamefont {D.}~\bibnamefont {Facoetti}},
  \bibinfo {author} {\bibfnamefont {M.}~\bibnamefont {Schir\'o}}, \bibinfo
  {author} {\bibfnamefont {M.}~\bibnamefont {Tarzia}},\ and\ \bibinfo {author}
  {\bibfnamefont {P.}~\bibnamefont {Vivo}},\ }\href
  {https://doi.org/10.1103/PhysRevB.103.014204} {\bibfield  {journal} {\bibinfo
   {journal} {Phys. Rev. B}\ }\textbf {\bibinfo {volume} {103}},\ \bibinfo
  {pages} {014204} (\bibinfo {year} {2021})}\BibitemShut {NoStop}%
\bibitem [{\citenamefont {Scoquart}\ \emph {et~al.}(2024)\citenamefont
  {Scoquart}, \citenamefont {Gornyi},\ and\ \citenamefont
  {Mirlin}}]{Scoquart2024Role}%
  \BibitemOpen
  \bibfield  {author} {\bibinfo {author} {\bibfnamefont {T.}~\bibnamefont
  {Scoquart}}, \bibinfo {author} {\bibfnamefont {I.~V.}\ \bibnamefont
  {Gornyi}},\ and\ \bibinfo {author} {\bibfnamefont {A.~D.}\ \bibnamefont
  {Mirlin}},\ }\href {https://doi.org/10.1103/PhysRevB.109.214203} {\bibfield
  {journal} {\bibinfo  {journal} {Phys. Rev. B}\ }\textbf {\bibinfo {volume}
  {109}},\ \bibinfo {pages} {214203} (\bibinfo {year} {2024})}\BibitemShut
  {NoStop}%
\bibitem [{\citenamefont {Pietracaprina}\ \emph {et~al.}(2016)\citenamefont
  {Pietracaprina}, \citenamefont {Ros},\ and\ \citenamefont
  {Scardicchio}}]{pietracaprina2016forward}%
  \BibitemOpen
  \bibfield  {author} {\bibinfo {author} {\bibfnamefont {F.}~\bibnamefont
  {Pietracaprina}}, \bibinfo {author} {\bibfnamefont {V.}~\bibnamefont {Ros}},\
  and\ \bibinfo {author} {\bibfnamefont {A.}~\bibnamefont {Scardicchio}},\
  }\href {https://doi.org/10.1103/PhysRevB.93.054201} {\bibfield  {journal}
  {\bibinfo  {journal} {Phys. Rev. B}\ }\textbf {\bibinfo {volume} {93}},\
  \bibinfo {pages} {054201} (\bibinfo {year} {2016})}\BibitemShut {NoStop}%
\bibitem [{\citenamefont {Abou-Chacra}\ \emph {et~al.}(1973)\citenamefont
  {Abou-Chacra}, \citenamefont {Thouless},\ and\ \citenamefont
  {Anderson}}]{abou1973selfconsistent}%
  \BibitemOpen
  \bibfield  {author} {\bibinfo {author} {\bibfnamefont {R.}~\bibnamefont
  {Abou-Chacra}}, \bibinfo {author} {\bibfnamefont {D.~J.}\ \bibnamefont
  {Thouless}},\ and\ \bibinfo {author} {\bibfnamefont {P.~W.}\ \bibnamefont
  {Anderson}},\ }\href {https://doi.org/10.1088/0022-3719/6/10/009} {\bibfield
  {journal} {\bibinfo  {journal} {J. Phys. C}\ }\textbf {\bibinfo {volume}
  {6}},\ \bibinfo {pages} {1734} (\bibinfo {year} {1973})}\BibitemShut
  {NoStop}%
\bibitem [{\citenamefont {Pietracaprina}\ \emph {et~al.}(2018)\citenamefont
  {Pietracaprina}, \citenamefont {Macé}, \citenamefont {Luitz},\ and\
  \citenamefont {Alet}}]{Pietracaprina2018Shift}%
  \BibitemOpen
  \bibfield  {author} {\bibinfo {author} {\bibfnamefont {F.}~\bibnamefont
  {Pietracaprina}}, \bibinfo {author} {\bibfnamefont {N.}~\bibnamefont
  {Macé}}, \bibinfo {author} {\bibfnamefont {D.~J.}\ \bibnamefont {Luitz}},\
  and\ \bibinfo {author} {\bibfnamefont {F.}~\bibnamefont {Alet}},\ }\href
  {https://doi.org/10.21468/SciPostPhys.5.5.045} {\bibfield  {journal}
  {\bibinfo  {journal} {SciPost Phys.}\ }\textbf {\bibinfo {volume} {5}},\
  \bibinfo {pages} {045} (\bibinfo {year} {2018})}\BibitemShut {NoStop}%
\bibitem [{\citenamefont {Atas}\ \emph {et~al.}(2013)\citenamefont {Atas},
  \citenamefont {Bogomolny}, \citenamefont {Giraud},\ and\ \citenamefont
  {Roux}}]{Atas2013Distribution}%
  \BibitemOpen
  \bibfield  {author} {\bibinfo {author} {\bibfnamefont {Y.~Y.}\ \bibnamefont
  {Atas}}, \bibinfo {author} {\bibfnamefont {E.}~\bibnamefont {Bogomolny}},
  \bibinfo {author} {\bibfnamefont {O.}~\bibnamefont {Giraud}},\ and\ \bibinfo
  {author} {\bibfnamefont {G.}~\bibnamefont {Roux}},\ }\href
  {https://doi.org/10.1103/PhysRevLett.110.084101} {\bibfield  {journal}
  {\bibinfo  {journal} {Phys. Rev. Lett.}\ }\textbf {\bibinfo {volume} {110}},\
  \bibinfo {pages} {084101} (\bibinfo {year} {2013})}\BibitemShut {NoStop}%
\bibitem [{\citenamefont {Cardy}(1988)}]{Cardy1988}%
  \BibitemOpen
  \bibinfo {editor} {\bibfnamefont {J.}~\bibnamefont {Cardy}},\ ed.,\
  \href@noop {} {\emph {\bibinfo {title} {{Finite-Size Scaling}}}},\ \bibinfo
  {number} {2}\ (\bibinfo  {publisher} {North Holland},\ \bibinfo {year}
  {1988})\BibitemShut {NoStop}%
\bibitem [{\citenamefont {Cardy}(1996)}]{Cardy_1996}%
  \BibitemOpen
  \bibfield  {author} {\bibinfo {author} {\bibfnamefont {J.}~\bibnamefont
  {Cardy}},\ }\href@noop {} {\emph {\bibinfo {title} {{Scaling and
  Renormalization in Statistical Physics}}}},\ Cambridge Lecture Notes in
  Physics\ (\bibinfo  {publisher} {Cambridge University Press},\ \bibinfo
  {year} {1996})\BibitemShut {NoStop}%
\bibitem [{Sup()}]{SupplMat}%
  \BibitemOpen
  \href@noop {} {}\bibinfo {howpublished} {See Supplemental Material for more
  data on the $r$-parameter, and a finite-size scaling analysis of the
  transition.}\BibitemShut {Stop}%
\bibitem [{\citenamefont {Pino}\ \emph {et~al.}(2017)\citenamefont {Pino},
  \citenamefont {Kravtsov}, \citenamefont {Altshuler},\ and\ \citenamefont
  {Ioffe}}]{Pino2017Multifractal}%
  \BibitemOpen
  \bibfield  {author} {\bibinfo {author} {\bibfnamefont {M.}~\bibnamefont
  {Pino}}, \bibinfo {author} {\bibfnamefont {V.~E.}\ \bibnamefont {Kravtsov}},
  \bibinfo {author} {\bibfnamefont {B.~L.}\ \bibnamefont {Altshuler}},\ and\
  \bibinfo {author} {\bibfnamefont {L.~B.}\ \bibnamefont {Ioffe}},\ }\href
  {https://doi.org/10.1103/PhysRevB.96.214205} {\bibfield  {journal} {\bibinfo
  {journal} {Phys. Rev. B}\ }\textbf {\bibinfo {volume} {96}},\ \bibinfo
  {pages} {214205} (\bibinfo {year} {2017})}\BibitemShut {NoStop}%
\bibitem [{Note1()}]{Note1}%
  \BibitemOpen
  \bibinfo {note} {This behavior can be deduced from the data reported in
  Ref.~\cite {kutlin2024investigating}, even if the fact is not explicitly
  commented.}\BibitemShut {Stop}%
\bibitem [{\citenamefont {Kac}(1956)}]{Kac1956Foundations}%
  \BibitemOpen
  \bibfield  {author} {\bibinfo {author} {\bibfnamefont {M.}~\bibnamefont
  {Kac}},\ }\bibinfo {title} {{Foundations of Kinetic Theory}},\ in\ \href
  {https://doi.org/doi:10.1525/9780520350694-012} {\emph {\bibinfo {booktitle}
  {Volume 3 Proceedings of the Third Berkeley Symposium on Mathematical
  Statistics and Probability, Volume III}}},\ \bibinfo {editor} {edited by\
  \bibinfo {editor} {\bibfnamefont {J.}~\bibnamefont {Neyman}}}\ (\bibinfo
  {publisher} {University of California Press},\ \bibinfo {address}
  {Berkeley},\ \bibinfo {year} {1956})\ pp.\ \bibinfo {pages}
  {173--200}\BibitemShut {NoStop}%
\bibitem [{\citenamefont {Campa}\ \emph {et~al.}(2014)\citenamefont {Campa},
  \citenamefont {Dauxois}, \citenamefont {Fanelli},\ and\ \citenamefont
  {Ruffo}}]{Campa2014Physics}%
  \BibitemOpen
  \bibfield  {author} {\bibinfo {author} {\bibfnamefont {A.}~\bibnamefont
  {Campa}}, \bibinfo {author} {\bibfnamefont {T.}~\bibnamefont {Dauxois}},
  \bibinfo {author} {\bibfnamefont {D.}~\bibnamefont {Fanelli}},\ and\ \bibinfo
  {author} {\bibfnamefont {S.}~\bibnamefont {Ruffo}},\ }\href
  {https://doi.org/10.1093/acprof:oso/9780199581931.001.0001} {\emph {\bibinfo
  {title} {{Physics of Long-Range Interacting Systems}}}}\ (\bibinfo
  {publisher} {Oxford University Press},\ \bibinfo {year} {2014})\BibitemShut
  {NoStop}%
\bibitem [{Note2()}]{Note2}%
  \BibitemOpen
  \bibinfo {note} {Notice that $\epsilon _0 = \pm \protect \sqrt {\log (2)}$ is
  the ground-state energy of the classical REM, that approximates well the edge
  of the spectrum $E_{\protect \mathrm {max}}$. The value of $\Gamma _c$ has
  been obtained using Eq. (35) of Ref.~\cite
  {Baldwin2016ManyBody}.}\BibitemShut {Stop}%
\bibitem [{\citenamefont {Faoro}\ \emph {et~al.}(2019)\citenamefont {Faoro},
  \citenamefont {Feigel’man},\ and\ \citenamefont {Ioffe}}]{FAORO2019167916}%
  \BibitemOpen
  \bibfield  {author} {\bibinfo {author} {\bibfnamefont {L.}~\bibnamefont
  {Faoro}}, \bibinfo {author} {\bibfnamefont {M.~V.}\ \bibnamefont
  {Feigel’man}},\ and\ \bibinfo {author} {\bibfnamefont {L.}~\bibnamefont
  {Ioffe}},\ }\href {https://doi.org/https://doi.org/10.1016/j.aop.2019.167916}
  {\bibfield  {journal} {\bibinfo  {journal} {Annals of Physics}\ }\textbf
  {\bibinfo {volume} {409}},\ \bibinfo {pages} {167916} (\bibinfo {year}
  {2019})}\BibitemShut {NoStop}%
\bibitem [{\citenamefont {Virtanen}\ \emph {et~al.}(2020)\citenamefont
  {Virtanen}, \citenamefont {Gommers}, \citenamefont {Oliphant} \emph
  {et~al.}}]{SciPy}%
  \BibitemOpen
  \bibfield  {author} {\bibinfo {author} {\bibfnamefont {P.}~\bibnamefont
  {Virtanen}}, \bibinfo {author} {\bibfnamefont {R.}~\bibnamefont {Gommers}},
  \bibinfo {author} {\bibfnamefont {T.~E.}\ \bibnamefont {Oliphant}}, \emph
  {et~al.},\ }\href {https://doi.org/10.1038/s41592-019-0686-2} {\bibfield
  {journal} {\bibinfo  {journal} {Nat. Meth.}\ }\textbf {\bibinfo {volume}
  {17}},\ \bibinfo {pages} {261} (\bibinfo {year} {2020})}\BibitemShut
  {NoStop}%
\bibitem [{\citenamefont {Balay}\ \emph {et~al.}(2024)\citenamefont {Balay},
  \citenamefont {Abhyankar}, \citenamefont {Adams} \emph {et~al.}}]{PETSc}%
  \BibitemOpen
  \bibfield  {author} {\bibinfo {author} {\bibfnamefont {S.}~\bibnamefont
  {Balay}}, \bibinfo {author} {\bibfnamefont {S.}~\bibnamefont {Abhyankar}},
  \bibinfo {author} {\bibfnamefont {M.~F.}\ \bibnamefont {Adams}}, \emph
  {et~al.},\ }\href@noop {} {\bibinfo {title} {{{PETS}c {W}eb page}}},\
  \bibinfo {howpublished} {\url{https://petsc.org/}} (\bibinfo {year}
  {2024})\BibitemShut {NoStop}%
\bibitem [{\citenamefont {Hernandez}\ \emph {et~al.}(2005)\citenamefont
  {Hernandez}, \citenamefont {Roman},\ and\ \citenamefont {Vidal}}]{SLEPc}%
  \BibitemOpen
  \bibfield  {author} {\bibinfo {author} {\bibfnamefont {V.}~\bibnamefont
  {Hernandez}}, \bibinfo {author} {\bibfnamefont {J.~E.}\ \bibnamefont
  {Roman}},\ and\ \bibinfo {author} {\bibfnamefont {V.}~\bibnamefont {Vidal}},\
  }\href {https://doi.org/10.1145/1089014.1089019} {\bibfield  {journal}
  {\bibinfo  {journal} {ACM Trans. Math. Softw.}\ }\textbf {\bibinfo {volume}
  {31}},\ \bibinfo {pages} {351–362} (\bibinfo {year} {2005})}\BibitemShut
  {NoStop}%
\end{thebibliography}%


\newpage
\onecolumngrid
\section*{Gap ratio analysis}
\label{app:sec:QREM_eps}

In the main text we provided numerical evidence for the absence of a localization transition in the QREM at the center of the spectrum and its presence at finite energy density, through the analysis of the beta function of the fractal dimension at $E=0$ and $E=E_{\text{max}}/4$. Additional evidence in support of this scenario can be found from the rescaled $r$-parameter $\phi$. 

\begin{figure}
    \centering
    \includegraphics[width=0.45\linewidth]{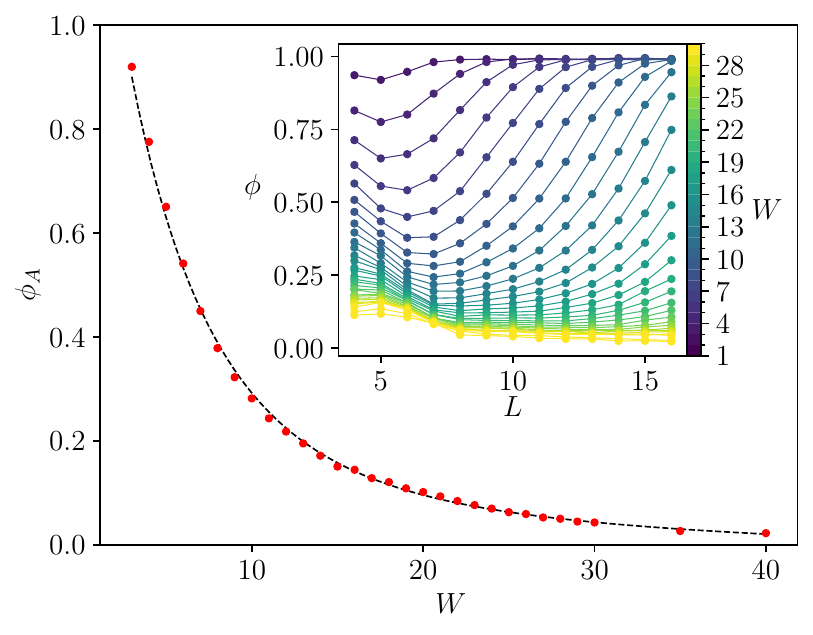}
    \caption{In the inset we show the scaling of $\phi$ with the system size, for different values of the disorder $W$, at $E=0$. In the main panel, the minima $\phi_A$ of the curves $\phi(L)$ are plotted as a function of the disorder parameter $W$. The points $\phi_A$ correspond to the turning points of the RG dynamics, as explained in the text. The points $\phi_A(W)$ are well fitted by a polynomial that extrapolates to 0 when $W \rightarrow \infty$.}
    \label{fig:phi_A}
\end{figure}

In the inset of Fig.~\ref{fig:phi_A} we plot $\phi$ as a function of system size at $E=0$. The minima of the curves $\phi(L)$, which we call $\phi_A$, are instead plotted as a function of the disorder strength $W$ in the main panel. These points correspond to turning points of the RG dynamics, i.e.\ the points at which the flow stops pointing towards the localized fixed point, and starts turning towards the ergodic one. Since $\phi_A$ decreases to 0 as the disorder $W$ increases, if a transition is present, then the turning points $\phi_A$ should extrapolate to the value $\phi_c = 0$ at a finite value of the disorder in the thermodynamic limit. The data in Fig.~\ref{fig:phi_A} instead fit well to a polynomial that extrapolates to 0 only in the infinite disorder limit, thus signaling that the system will eventually flow to the ergodic phase for every finite value of $W$. 

\section*{Finite-size scaling of the rescaled gap ratio}

Here we provide additional details regarding the finite-size scaling of our numerical data. By denoting with $A$ any of the spectral observables considered in the main text, its beta function is defined as 
\begin{equation}
    \beta = \frac{\rmd \ln A}{\rmd \ln N}.
\end{equation}
Under the one-parameter scaling (1PS) hypothesis, the beta function is a function of $A$ alone, i.e.\ $\beta \equiv \beta(A)$. The dependence of $A$ on the system size $\ln N$ can be obtained by integrating the equation defining the beta function, i.e.~$\beta=\rmd \ln A/ \rmd \ln N$, by separation of variables. Defining the primitive $G(A) = \int_{A_0}^A \rmd A'/ (A' \beta(A'))$, one gets
\begin{equation}
    A(N)=f(N/\tilde N_0),
\end{equation}
where $\tilde N_0 \equiv N_0 e^{-G(A_0)}$ and the function $f(x)=G^{-1}(\ln x)$. This scaling form can be modified by taking into account finite-size corrections from the \emph{irrelevant} part of the RG flow. Denoting by $\omega$ the scaling dimension of the least irrelevant operator, it can be shown that~\cite{Niedda2024}
\begin{equation}
    \label{eq:1PS_irrelevant_corrections}
    A(N) =f(N/\tilde{N}_0)+ f_1(N/\tilde{N}_0) N^{-\omega},
\end{equation}
with a second scaling function $f_1$. 

The scaling theory described above holds when the critical point is a well-isolated fixed point of the RG flow, as in the Anderson model in finite dimensions~\cite{altshuler2024renormalization}. When, instead, the critical value of the observable $A$ drifts to the localized value $A=0$, one direction of the RG flow must necessarily become flat and the scaling theory must be supplemented with the addition of a second parameter~\cite{vanoni2023renormalization,Niedda2024}.

\begin{figure}
    \centering
    \includegraphics[width=0.45\columnwidth]{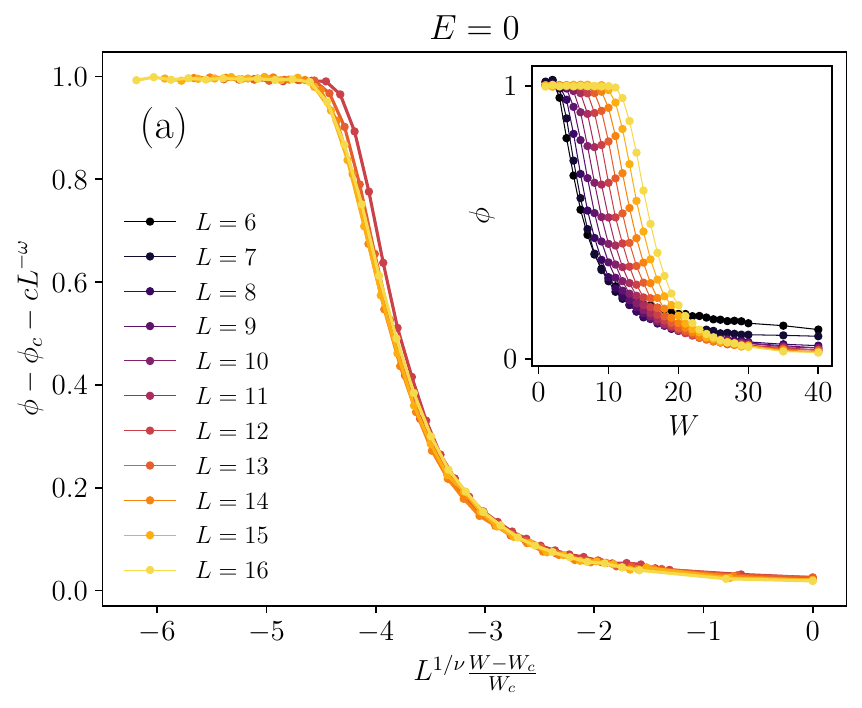}
    \includegraphics[width=0.45\columnwidth]{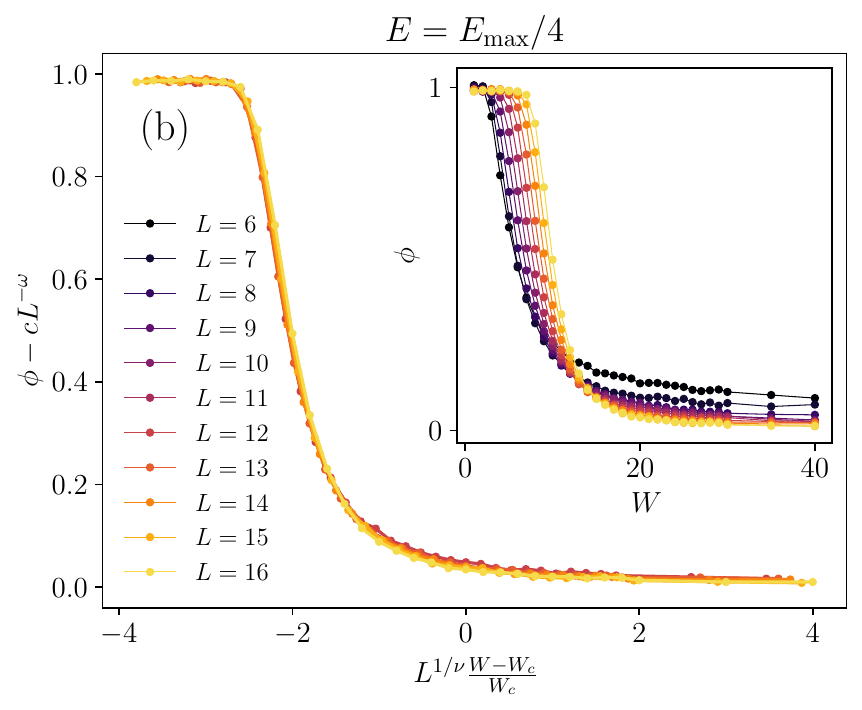}
    \caption{(a) Naive finite-size scaling for the rescaled $r$-parameter $\phi$ at $E=0$, where the model does not have a localization transition. Despite this known fact, by fixing $\phi_c=0$, it is still possible to obtain a good data collapse with a \emph{fake} critical point at $W_c = 40$, obtaining (fake) critical exponents $\nu = 1.5$ and $\omega = 2$. In the main plot we show the collapse for the larger sizes, $L>11$, while the inset contains the same data down to $L=6$, where finite-size effects are stronger.
    (b) Finite-size scaling for the average rescaled gap ratio $\phi$ at $E = E_{\mathrm{max}}/4$. The collapse is obtained for $W_c \simeq 20$ and $\phi_c = 0$. The critical exponent $\nu \simeq 2$ and $\omega = 2$. Despite the presence of a localization transition (as discussed in the main text), the one-parameter scaling ansatz does not apply, and therefore the critical properties extracted from the FSS should not be trusted.}
    \label{fig:scalingQREM}
\end{figure}

\begin{figure}
    \centering
    \includegraphics[height=0.38\columnwidth]{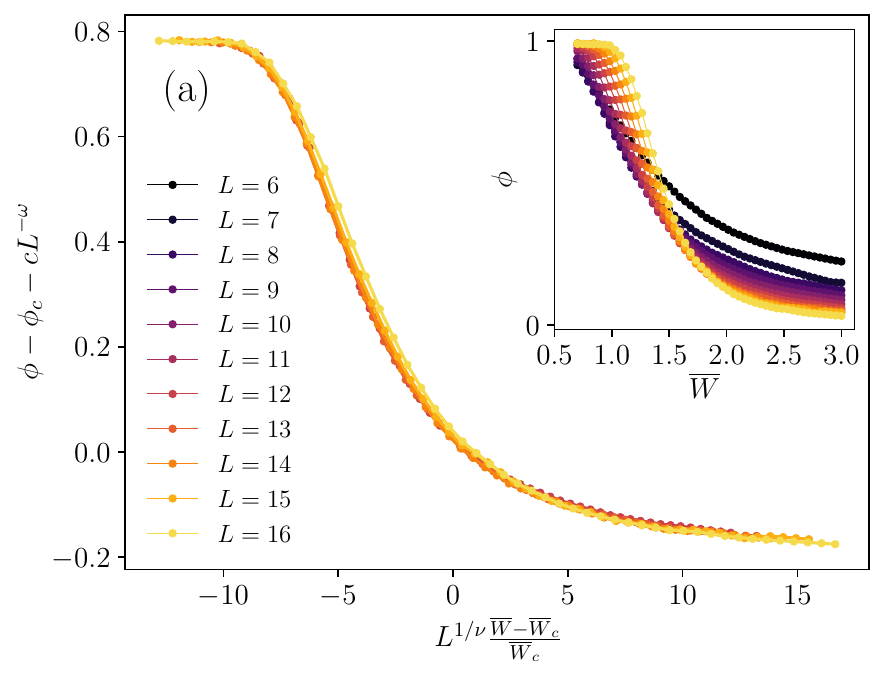}
    \includegraphics[height=0.38\columnwidth]{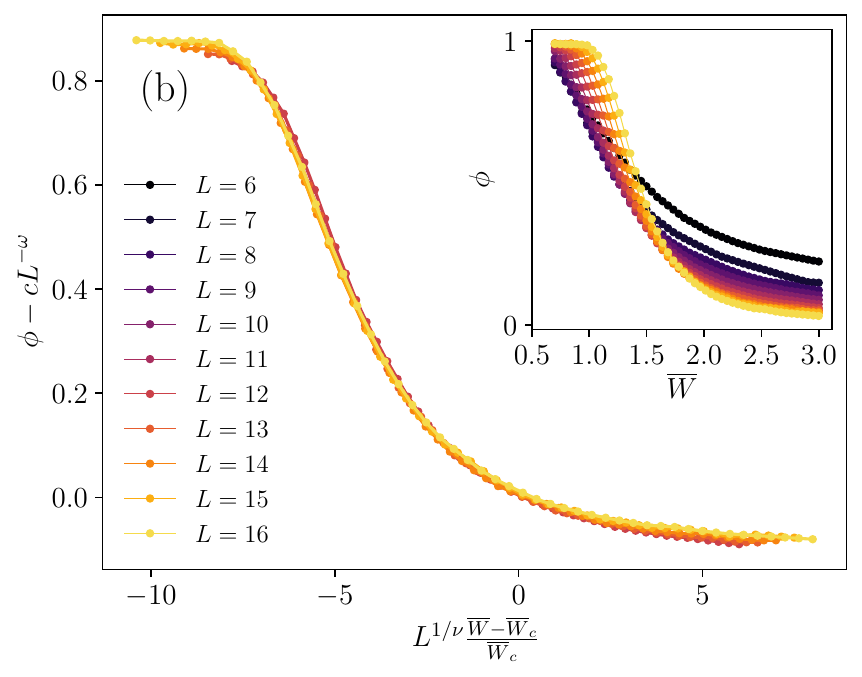}
    \caption{Finite-size scaling for the rescaled $r$-parameter $\phi$ at $E=0$ as a function of $\overline{W} = W/\sqrt{L}\log L$. (a) The collapse is obtained leaving both $\overline{W}_c$ and $\phi_c$ as free parameters, obtaining from the fit $\overline{W}_c \simeq 1.7$ and $\phi_c \simeq 0.2$. This is compatible with the result obtained from the zeros of the beta function (see inset of Fig.~\ref{fig:phi_A}). The critical exponents are found to be $\nu \simeq 0.9$ and $\omega = 2$. 
    (b) The collapse is obtained keeping $\phi_c = 0$ and $\overline{W}_c \simeq 2$, obtained in the main text from the zeros of beta function and the energies of the dynamical system. This is compatible with the result obtained from the zeros of the beta function (see inset of Fig.~\ref{fig:phi_A}). The critical exponent are $\nu \simeq 1$ and $\omega = 1$.}
    \label{fig:phi_FSS}
\end{figure}

If one blindly applies the 1PS ansatz \eqref{eq:1PS_irrelevant_corrections} to the QREM at infinite temperature, where there is no localization transition and the model is ergodic for all values of $W$, they will obtain an inconsistent result. Referring to Fig.~\ref{fig:scalingQREM}(a), one can see that a fairly good scaling collapse of the rescaled $r$-parameter $\phi$ is obtained for large system sizes, identifying a \emph{fake} critical point at $W_c \simeq 40$ with critical exponents $\nu \simeq 3/2$ and $\omega \simeq 2$. 

On the other hand, at finite energy $E=E_{\text{max}}/4$ the data are compatible with a localization transition at $W_c \simeq 20$. Moreover, from the beta function of the fractal dimension reported in the main text, it is expected that the transition be described by a 2PS theory. It is therefore wrong to use the 1PS form Eq.~\eqref{eq:1PS_irrelevant_corrections} to collapse the data. In Fig.~\ref{fig:scalingQREM}(b), we show the collapse of the rescaled $r$-parameter at $E=E_{\text{max}}/4$, obtained with critical exponents $\nu \simeq 2$ and $\omega=2$: the collapse yields a critical value $\phi_c \simeq 0.1$ at a disorder $W_c \simeq 20$, compatible with the value obtained by studying the turning points $\phi_A$.

Finally, we report two scaling collapses of $\phi$ obtained with the rescaling of the disorder $\overline{W}=W/\sqrt{L} \ln L$, which makes a localization transition emerge at the center of the spectrum. However, one can observe that, while the collapses look convincing to the naked eye, they lead to a misleading interpretation of the data. The analysis of the beta function of $\phi$ performed in the main text shows that the transition is described by a 2PS theory, with a drift of the critical point to $\phi_c=0$. If instead $\phi_c$ is left as a free parameter in the collapse, one obtains the incorrect value $\phi_c \simeq 0.2$ in Fig.~\ref{fig:phi_FSS}(a). On the other hand, a collapse in which $\phi_c=0$ is imposed is shown in Fig.~\ref{fig:phi_FSS}(b): despite being a good collapse, with $W_c \simeq 2$ as predicted by the zeros of the beta function, it gives the nonphysical result $\phi < 0$ for $\overline{W} > \overline{W}_c$, signaling that the 1PS ansatz is not valid for the QREM at $E=0$.

\end{document}